\documentclass[fleqn,usenatbib]{mnras}

\usepackage{newtxtext,newtxmath}
\usepackage{booktabs}

\usepackage[T1]{fontenc}

\DeclareRobustCommand{\VAN}[3]{#2}
\let\VANthebibliography\thebibliography
\def\thebibliography{\DeclareRobustCommand{\VAN}[3]{##3}\VANthebibliography}

\usepackage{graphicx}	
\usepackage{amsmath}	

\title[Testing cross-correlations to search for molecular features in JWST observations of transiting exoplanets]{Testing the performance of cross-correlation techniques to search for molecular features in JWST NIRSpec G395H observations of transiting exoplanets}

\author[E. Esparza-Borges et al.]{
Emma Esparza-Borges,$^{1,2}$\thanks{E-mail: emma.esparza.borges@gmail.com}
Mercedes L\'opez-Morales,$^{3}$
Enric Pall\'e,$^{1,2}$
Vladimir Makhnev,$^{4}$
Iouli Gordon,$^{4}$
\newauthor
Robert Hargreaves,$^{4}$
James Kirk,$^{5}$
Claudio C\'aceres,$^{6}$
Ian J.M. Crossfield,$^{7}$
Nicolas Crouzet,$^{8}$
Leen Decin,$^{9}$
\newauthor
Jean-Michel Désert,$^{10}$
Laura Flagg,$^{11}$
Antonio Garc\'ia Muñoz,$^{12}$
Joseph Harrington,$^{13}$
Karan Molaverdikhani,$^{14}$
\newauthor
Giuseppe Morello,$^{15,16}$
Nikolay Nikolov,$^{3}$
Arif Solmaz,$^{17}$
Benjamin V.\ Rackham$^{18,19}$
and Seth Redfield.$^{20}$
\\
$^{1}$ Instituto de Astrof\'isica de Canarias, E-38200 La Laguna, Tenerife, Spain\\
$^{2}$ Departamento de Astrof\'isica, Universidad de La Laguna, E-38206 La Laguna, Tenerife, Spain\\
$^{3}$ Space Telescope Science Institute, 3700 San Martin Drive, Baltimore MD 21218 USA\\
$^{4}$ Center for Astrophysics $\mid$ Harvard $\&$ Smithsonian, Atomic and Molecular Physics Division, Cambridge, MA 02138, USA\\
$^{5}$ Department of Physics, Imperial College London, Prince Consort Road,
London SW7 2AZ, UK\\
$^{6}$ Instituto de Astrofisica, Departamento de Fisica y Astronomia, Facultad de Ciencias Exactas, Universidad Andres Bello. Santiago, Chile\\
$^{7}$ Department of Physics and Astronomy, University of Kansas, Lawrence, KS, USA\\
$^{8}$ Kapteyn Astronomical Institute, Rijksuniversiteit Groningen, Postbus 800, 9700 AV Groningen, The Netherlands\\
$^{9}$ Institute of Astronomy, KU Leuven, Celestijnenlaan 200D, Belgium\\
$^{10}$ Anton Pannekoek Institute for Astronomy, University of Amsterdam, Amsterdam, The Netherlands\\
$^{11}$ Johns Hopkins University, Baltimore, MD 21218, USA\\
$^{12}$ Paris-Saclay, Université Paris Cité, CEA, CNRS, AIM, 91191 Gif-sur-Yvette, France\\
$^{13}$ Planetary Sciences Group, Department of Physics and Florida Space Institute, University of Central Florida, Orlando, FL, USA\\
$^{14}$ Universitäts-Sternwarte, Ludwig-Maximilians-Universität München, Scheinerstrasse 1, D-81679 München, Germany\\
$^{15}$ Instituto de Astrofísica de Andalucía (IAA-CSIC), Glorieta de la Astronomía s/n, 18008 Granada, Spain\\
$^{16}$ INAF - Osservatorio Astronomico di Palermo, Piazza del Parlamento, 1, 90134 Palermo, Italy\\
$^{17}$ İstanbul Health and Technology University, 34445, İstanbul, Türkiye\\
$^{18}$ Department of Earth, Atmospheric and Planetary Sciences, Massachusetts Institute of Technology, Cambridge, MA 02139, USA\\
$^{19}$ Kavli Institute 
for Astrophysics and Space Research, Massachusetts Institute of Technology, Cambridge, MA 02139, USA\\
$^{20}$ Astronomy Department and Van Vleck Observatory, Wesleyan University, Middletown, CT 06459, USA
}

\date{XXX. Received YYY; in original form ZZZ}

\pubyear{\the\year{}}

\begin{document}
\label{firstpage}
\pagerange{\pageref{firstpage}--\pageref{lastpage}}
\maketitle

\begin{abstract}
Cross-correlations techniques  offer an alternative method to search for molecular species in JWST observations of exoplanet atmospheres. In a previous article, we applied cross-correlation functions for the first time to JWST NIRSpec/G395H observations of exoplanet atmospheres, resulting in a detection of CO in the transmission spectrum of WASP-39b and a tentative detection of CO isotopologues. Here we present an improved version of our cross-correlation technique and an investigation into how efficient the technique is when searching for other molecules in JWST NIRSpec/G395H data. Our search results in the detection of more molecules via cross-correlations in the atmosphere of WASP-39b, including $\rm H_{2}O$ and $\rm CO_{2}$, and confirms the CO detection. This result proves that cross-correlations are a robust and computationally cheap alternative method to search for molecular species in transmission spectra observed with JWST. We also searched for other molecules ($\rm CH_{4}$, $\rm NH_{3}$, $\rm SO_{2}$, $\rm N_{2}O$, $\rm H_{2}S$, $\rm PH_{3}$, $\rm O_{3}$ and $\rm C_{2}H_{2}$) that were not detected, for which we provide the definition of their cross-correlation baselines for future searches of those molecules in other targets. We find that that the cross-correlation search of each molecule is more efficient over limited wavelength regions of the spectrum, where the signal for that molecule dominates over other molecules, than over broad wavelength ranges. In general we also find that Gaussian normalization is the most efficient normalization mode for the generation of the molecular templates.


\end{abstract}




\section{Introduction}

Before the launch of the James Webb Space Telescope (JWST),  bonafide detections of molecular species in  exoplanet atmospheres from space were mostly limited to $\rm H_2O$ \citep[e.g.,][]{Sing2016}. 
The limited wavelength coverage of available space-borne instrumentation at the time 
 prevented the detection of other molecules. 
JWST has changed that 
by providing for the first time access to a wider range of wavelengths, which are sensitive to a larger number of molecular species. Within three years, the catalog of molecules detected in exoplanet atmospheres using JWST has expanded significantly, with high-confidence detections of 
${\rm H_2O}$, ${\rm CO_2}$, ${\rm CO}$, ${\rm SO_2}$ \citep[e.g.,][]{JWSTERS2023,Rustamkulov2023,Feinstein2023,Alderson2023,EsparzaBorges23,Grant2023, Kirk2024}, and ${\rm CH_4}$ \citep[e.g.,][]{Bell2023}. Detections of more exotic molecules, such as ${\rm CS_2}$ \citep{Benneke2024, Holmberg2024}, 
have also been reported. 

A standard approach to identifying molecular species in exoplanet atmospheres is via retrieval models 
that, although computationally expensive, are able to estimate molecular abundances, atmospheric temperatures, and elemental ratios \citep[e.g.,][]{Banerjee24, Kirk25, Scmidt25, Kanumalla25, Basilicata25}. 
Cross-correlation functions \citep[CCFs; e.g.,][]{Snellen2010_CC,Brogi2012, Rodler2012} offer another approach, which is arguably less efficient than  retrievals at constraining atmospheric parameters, but more time efficient at detecting molecular features and less sensitive to systematics such as bumps and offsets that appear in transmission spectra and that are often attributed to instrumental, limb-darkening effects, etc, and vary between pipeline reductions and observing epochs \citep[e.g.,][]{May2023}. Additionally, high resolution observations increase the sensitivity to molecular signals, and can separate the planet's signal from the stellar signal via Doppler shifts \citep{Snellen15}.
CCFs are regularly used in very high-resolution observations from the ground \citep[e.g.,][]{Brogi16,Hoeijmakers19,Hoeijmakers20,Yan19,Stangret20,Tabernero21,Carleo22,Nortmann25}, but have now also been demonstrated to work on the highest resolution modes of JWST, specifically in transmission and direct imaging observations of exoplanet atmospheres with NIRSpec G395H \citep{EsparzaBorges23, Gandhi23}. 

The CCFs detection of CO in the atmosphere of WASP-39b \citep[][hereafter EB23]{EsparzaBorges23},  was successful as a pilot program with a fairly simplified approach for a targeted molecule, a narrowly limited wavelength range, and normalizing the data using a simple polynomial function. In this paper, we present the results of additional work undertaken 1) to examine whether detections of CO in JWST NIRSpec G395H observations can be improved by expanding the approach adopted in EB23, and 2) to test the performance of the technique when searching for a number of other molecules expected to be stable in the atmospheres of hot gas giants like WASP-39b. 

Section~\ref{sec: methodology} briefly summarizes the methodology used in EB23, and describes how  that methodology has been expanded and generalized to be applied to other molecules. Section~\ref{sec: ccresults} shows the results when applying the new methods to the observed JWST NIRSpec/G395H transmission spectrum of WASP-39b. Our findings are summarized and discussed in Section~\ref{sec:conclusions}.

\section {Methodology}\label{sec: methodology}

In EB23, we presented a CCF method to search for CO in the observed transmission spectrum of WASP-39b with JWST NIRSpec/G395H (ERS-1366; PI: Batalha). Briefly, this method consisted in extracting the NRS2 detector transmission spectrum of WASP-39b at the detector pixel level and cross-correlating that transmission spectrum with a pure CO template generated using the public radiative transfer code \texttt{petitRADTRANS} \citep{MolliereSnellen2019}, using the known parameters of the planet \citep{Faedi11,Mancini18}. The cross-correlation was only applied to a portion of the JWST NIRSpec/G395H NRS2 spectrum between 4.6 and 5.0 $\mu$m, where the CO signal was expected to dominate over noise and over additional potential contributions from other molecules (see Fig.~1 in EB23). Prior to performing the cross-correlation search, both the observed and template spectra were normalized between those wavelengths using a fourth-order polynomial function. The peak of the resultant CCF -- expected to appear shifted in velocity to a value consistent with the relative velocity of the system with respect to JWST at the time of the observations -- was then normalized using the average of the CCF values over a range of Doppler velocities with low CCF values (see Fig.~3 in EB23). We then reported a CCF signal-to-noise ratio (CCF-SNR), defined as the ratio between each value of the CCF and the average value over that region (see Figures~\ref{fig:Baselines}, \ref{fig:Freq_Baselines}, \ref{fig:Freq_Baselines_nondetected_NRS1}, \ref{fig:Freq_Baselines_nondetected_NRS2} and Section 4 in EB23). 

Here we describe our refined approach to search for CO, and extend the method to 
other molecules expected to 1) be stable at temperature and pressure conditions typical of hot, gas giant exoplanets, and 2) have absorption features in the wavelength range covered by JWST NIRSpec G395H (i.e., 2.71 - 3.72 $\rm \mu m$ for NRS1 detector and 3.82-5.18 $\rm \mu m$ for NRS2 detector), based on the publicly available linelists HITEMP \citep{Rothman10} and HITRAN \citep{Gordon22}. 
The molecules included in our study are summarized in Table~\ref{tab: linelists}, together with references to the source of their linelists. The individual line-intensities of each molecule over the wavelength range covered by JWST NIRSpec/G395H are shown in Appendix Figure~\ref{fig:cross-sections_molecules}. In the following subsections, we describe our refined analysis process, highlighting the parts that are similar to, or differ from, the process followed in EB23.

\subsection{Transmission Spectrum}
\begin{figure*}
\includegraphics[width=\textwidth]{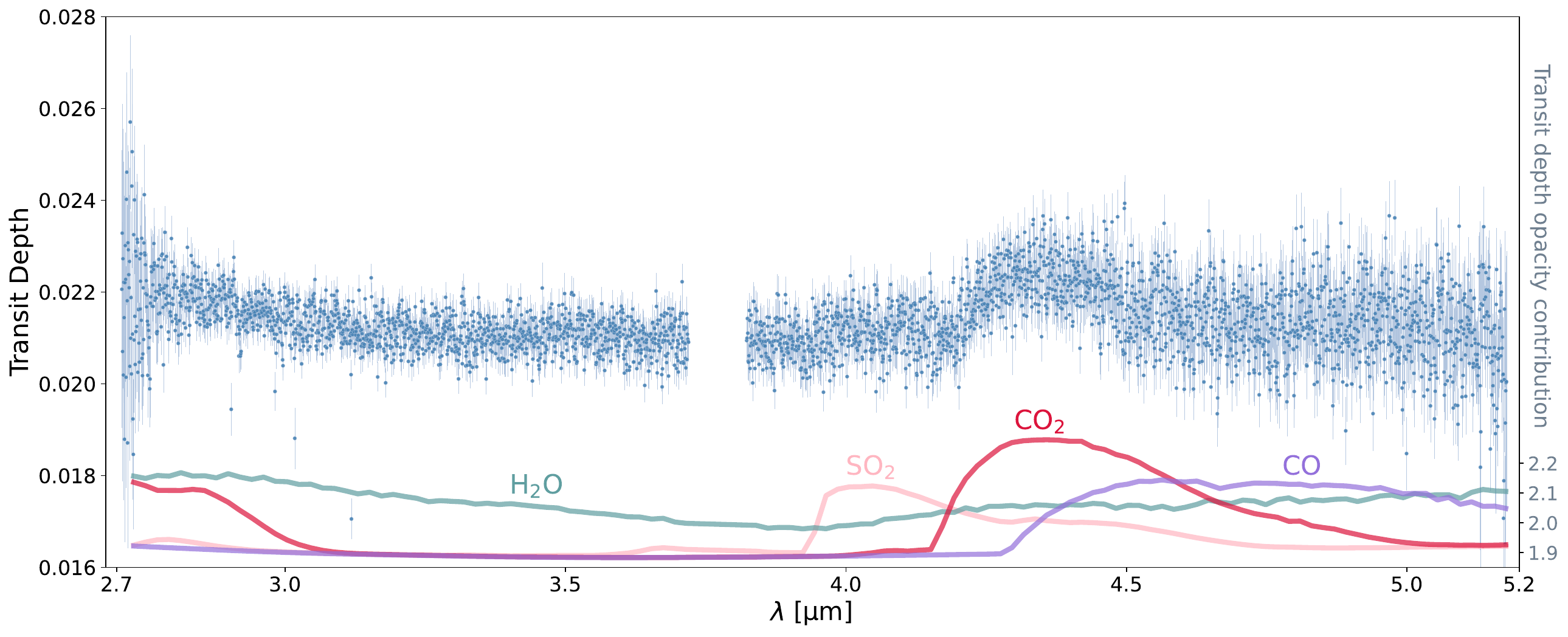}
\caption{Transmission spectrum of WASP-39b obtained from the JTEC ERS Program NIRSpec G395H observations across NRS1 and NRS2 detectors extracted at pixel-level by the \texttt{Tiberius} pipeline. The right axis shows the relative opacity contributions of $\rm H_{2}O$, $\rm SO_{2}$, $\rm CO_{2}$ and $\rm CO$, which are expected to be the dominant spectral features in this spectral range in WASP-39b. 
}
\label{fig:TS}
\end{figure*}

To test our technique, we use the same JWST ERS-1366 NIRSpec/G395H NRS2 spectrum of WASP-39b presented in EB23. Expanding from EB23, we also use the 
JWST ERS-1366 NIRSpec/G395H NRS1 spectrum of WASP-39b extracted using
the \texttt{Tiberius} pipeline \citep{Kirk17,Kirk21} in the same manner as described in EB23. The extracted pixel level transmission spectra, with mean resolving powers of $\rm R \sim 2380$ for NRS1 and $\rm R \sim 3454$ for NRS2 are shown in Figure~\ref{fig:TS}. The figure also shows the cross-sections for the molecules detected in that spectrum by previous analyses \citep{Rustamkulov2023,Feinstein2023,Alderson2023,EsparzaBorges23,Grant2023}.

\begin{table}
\caption{Molecules searched in this study and details of their linelists.}
\begin{tabular}{lcc}
      \toprule \toprule
      
      Molecule & Linelist name & Linelist origin \\ 
      \midrule
      
      $\rm H_{2}O$ & `$\rm H2O\_main\_iso$' &  HITEMP $^{(1)}$ \\
      $\rm CO_{2}$ & `$\rm CO2\_main\_iso$' &  HITEMP $^{(1)}$ \\
      $\rm CO$ & `$\rm CO\_all\_iso$' &  HITEMP $^{(2)}$ \\
      $\rm CH_{4}$ & `$\rm CH4\_main\_iso$' &  HITEMP $^{(3)}$\\
      $\rm NH_{3}$ & `$\rm NH3\_main\_iso$' &  HITRAN $^{(4)}$\\
      $\rm SO_{2}$ & `$\rm SO2\_HITRAN20\_pRT$' &  HITRAN $^{(4)}$\\
      $\rm N_{2}O$ & `$\rm N2O\_HAPI2\_P$' &  HITEMP $^{(5)}$ \\
      $\rm H_{2}S$ & `$\rm H2S\_main\_iso$' &  HITRAN $^{(4)}$ \\
      
      
      $\rm PH_{3}$ & `$\rm PH3\_main\_iso$' &  HITRAN $^{(4)}$ \\
      $\rm O_{3}$ & `$\rm O3\_main\_iso$' &  HITRAN $^{(4)}$ \\
      $\rm C_{2}H_{2}$ & `$\rm 12C2-1H2$' &  HITRAN $^{(4)}$\\
      
      \bottomrule \bottomrule
      
      
          
      \end{tabular}
\begin{tabular}{l}
      
      $^{(1)}$ HITEMP \citep{Rothman10}.\\
      $^{(2)}$ HITEMP update for CO \citep{Li15}\\
      $^{(3)}$ HITEMP update for $\rm CH_{4}$ \citep{Hargreaves20}\\
      $^{(4)}$ HITRAN2020 \citep{Gordon22}.\\
      $^{(5)}$ HITEMP update for $\rm N_{2}O$ \citep{Hargreaves19}\\
      \end{tabular}
      
      \label{tab: linelists}
\end{table}

\subsection{Generation of Templates}
We generated template transmission spectra for each molecule across the wavelengths covered by JWST NIRSpec/G395H, using the radiative transfer code \texttt{petitRADTRANS} \citep{MolliereSnellen2019}. For the simulated planet, we adopted the same stellar and planetary parameters used in the EB23 \citep{Faedi11}. We used linelists of all the molecules listed in Table 1 from HITEMP \citep{Rothman10,Li15,Hargreaves19,Hargreaves20} when available, using HITRAN2020 \citep{Gordon22} linelists otherwise, and assumed atmospheres composed purely of each molecule of interest. As in EB23, we first calculated each template transmission spectrum using the line-by-line radiative transfer mode of \texttt{petitRADTRANS} with resolving power  $\rm \lambda/\Delta\lambda = 10^{6}$, and then  resampled each spectrum  with \texttt{spectres} \citep{spectres} to match the resolution of the observations. 

An updated HITEMP line list of CO$_2$ has since become available \citep{Hargreaves25}, this line list does not affect the conclusions of this work, but is recommended for use in future investigations due to the increased accuracy and completeness.

\subsection {Generalization of the spectra normalization step}



\begin{figure*}
\includegraphics[width=\textwidth]{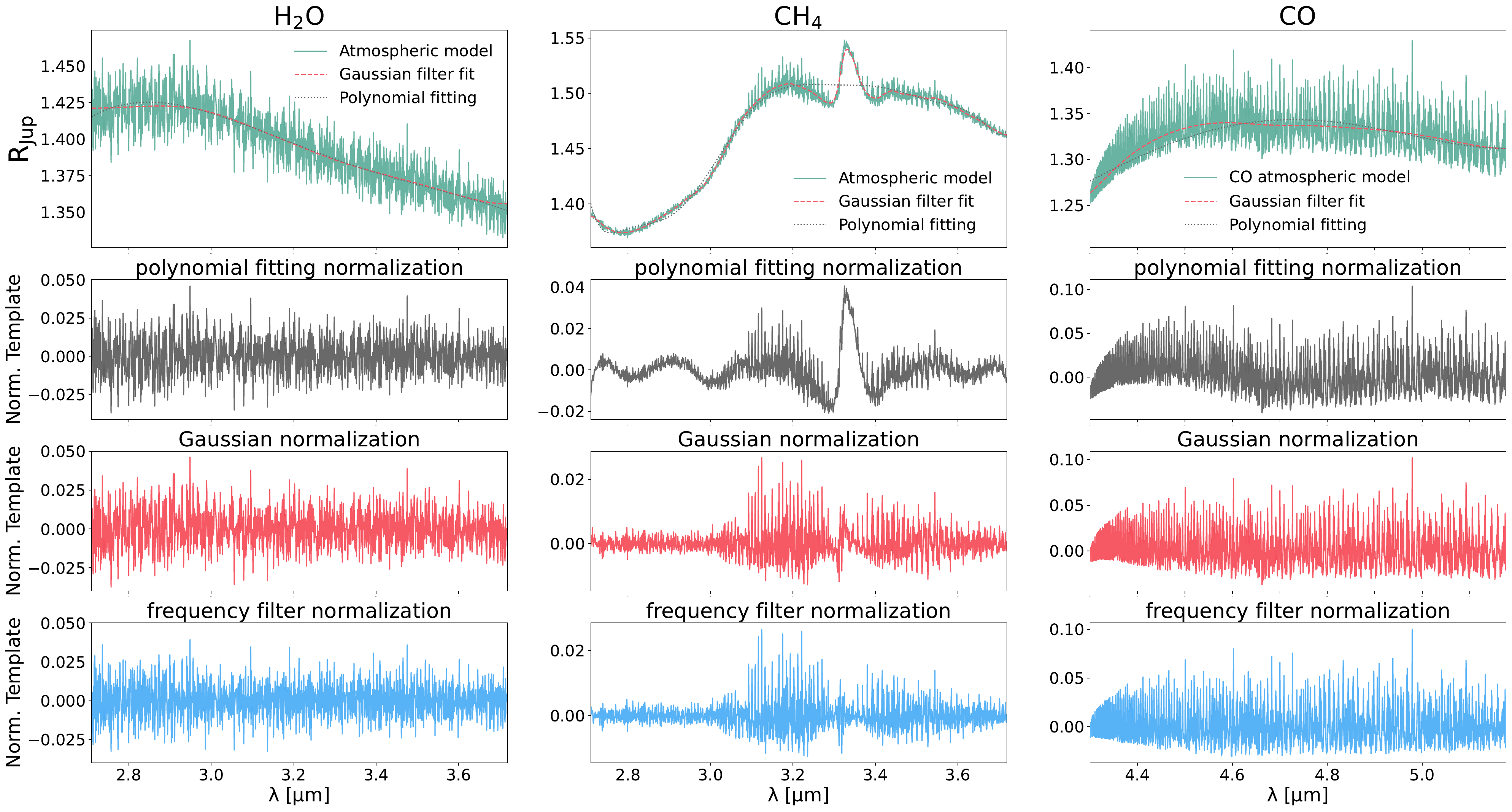}
\caption{Examples of the template extraction from the atmospheric models at JWST's spectral resolution (top panels) for three molecules, $\rm H_{2}O$ (left column panels), $\rm CH_{4}$ (middle column panels) and $\rm CO$ (right column panels), using three different normalization approaches. The second, third, and final rows show the resulting normalized templates using the polynomial fitting, Gaussian, and high-pass frequency filtering approaches, respectively. 
}
\label{fig:Norm_examples}
\end{figure*}

As an improvement from the EB23 analysis, we generalize the normalization step by performing three different normalization approaches and selecting {\it a posteriori} the most suitable approach for each molecule. The three normalization approaches implemented are: polynomial fitting (as in EB23), one-dimensional Gaussian filter fitting \citep[frequently used in high-resolution cross-correlation studies, e.g.,][]{Yang_Chen24}, and frequency filtering with a Butterworth high-pass filter \citep{Butterworth1930}. To illustrate the effectiveness of each approach works, we show in Figure~\ref{fig:Norm_examples} examples for ${\rm H_2O}$, ${\rm CH_4}$ and ${\rm CO}$.

For the 1-D Gaussian filter normalization approach we used the \texttt{gaussian$\rm\_$filter1d} function in \texttt{SciPy.ndimage} \citep{2020SciPy} to fit the spectrum, setting the standard deviation for the Gaussian kernel at $\rm \sigma = 100$. For the frequency filtering approach, we used the function \texttt{butter} in \texttt{SciPy.signal} \citep{2020SciPy} to compute the coefficients of a fifth-order Butterworth high-pass filter, setting the critical cutoff frequency at $\rm Wn=500$ and using a sampling frequency of $\rm fs=10000$. We used the function \texttt{sosfiltfilt} in \texttt{SciPy.signal} to apply the filter to the spectrum. For a description of the approach based on polynomial fitting, see EB23.

The performance of the normalization approaches depends on the shape and spectral features of the molecular absorption, i.e., slopes/trends or Gaussian-like/broad spectral features. To illustrate the performance of each normalization approach, we show in Figure~\ref{fig:Norm_examples} the results of normalizing the $\rm H_{2}O$ and $\rm CH_{4}$ bands in the NRS1 spectrum and the CO band in NRS2 with each normalization approach. 
This example shows how the $\rm H_{2}O$ and $\rm CH_{4}$ bands require different normalization approaches to extract a normalized molecular template suitable for cross-correlations. Thus, in the context of a systematic search of different molecules, it is necessary to use different normalization approaches and select the one that works best for each molecule. In general, we find that 1D Gaussian filter fitting works best for the molecules detected in this study, as described in more detail in Section~\ref{sec: ccresults}. In addition, we find that the polynomial fitting normalization is the least efficient normalization mode, as it might introduce artifacts in the generated templates for being less suitable to fit molecules presenting broad spectral features or curved trends. For this reason, we discarded the polynomial fitting normalization in this study.



\subsection{Cross-correlation analysis}\label{subsec: ccanalysis}

The next step of our analysis consists of cross-correlating the normalized pixel-level spectrum of WASP-39b with the 
normalized template of each molecule. We followed the procedure described in EB23, which we briefly summarize here for the reader's convenience.

We used the \texttt{crosscorrRV} algorithm in \texttt{PyAstronomy} \citep{PyAstronomy} to perform cross-correlations. This function Doppler-shifts the template over a range of RV offsets, evaluates the shifted template using linear interpolation to match the wavelength points of the observed spectrum and calculates the cross-correlation function at each RV shift as: 

\begin{equation}
    \rm CC(RV) = \sum\limits_{\lambda} S(\lambda) \times T(\lambda - \Delta_{\lambda,RV})
\end{equation}
where $\rm \Delta_{\lambda,RV}$ is the Doppler shift implemented as $\rm \Delta_{\lambda,RV} = \lambda(RV/c)$, $\rm T(\lambda - \Delta_{\lambda,RV})$ is the shifted template and $\rm S(\lambda)$ is the observed spectrum. Additionally, before computing the cross-correlation, we zero-padded the template to avoid edge artifacts in the cross-correlation.

Cross-correlations were performed over a range of radial velocities between $\rm \pm 1000~km~s^{-1}$, in $\rm 6~km~s^{-1}$ steps. As explained in Section 4 of EB23, $\rm 6~km~s^{-1}$ corresponds to a fraction of the RV pixel-to-pixel sampling. To test how these results depend on the resolution in RV space, we repeated this cross-correlation analysis for wider RV steps (20, 43 and 64 km/s). We present the results of these tests in Appendix~\ref{ap: rvsteps}. These results show consistent molecular detections, being the CCF maxima at the expected velocity (Figure~\ref{fig: app_RVsteps_CCFs}) and without substantial variations in the detection significances when varying the RV step in the cross-correlation (Figure~\ref{fig: app_RVsteps}).

\begin{figure*}
\includegraphics[width=\textwidth]{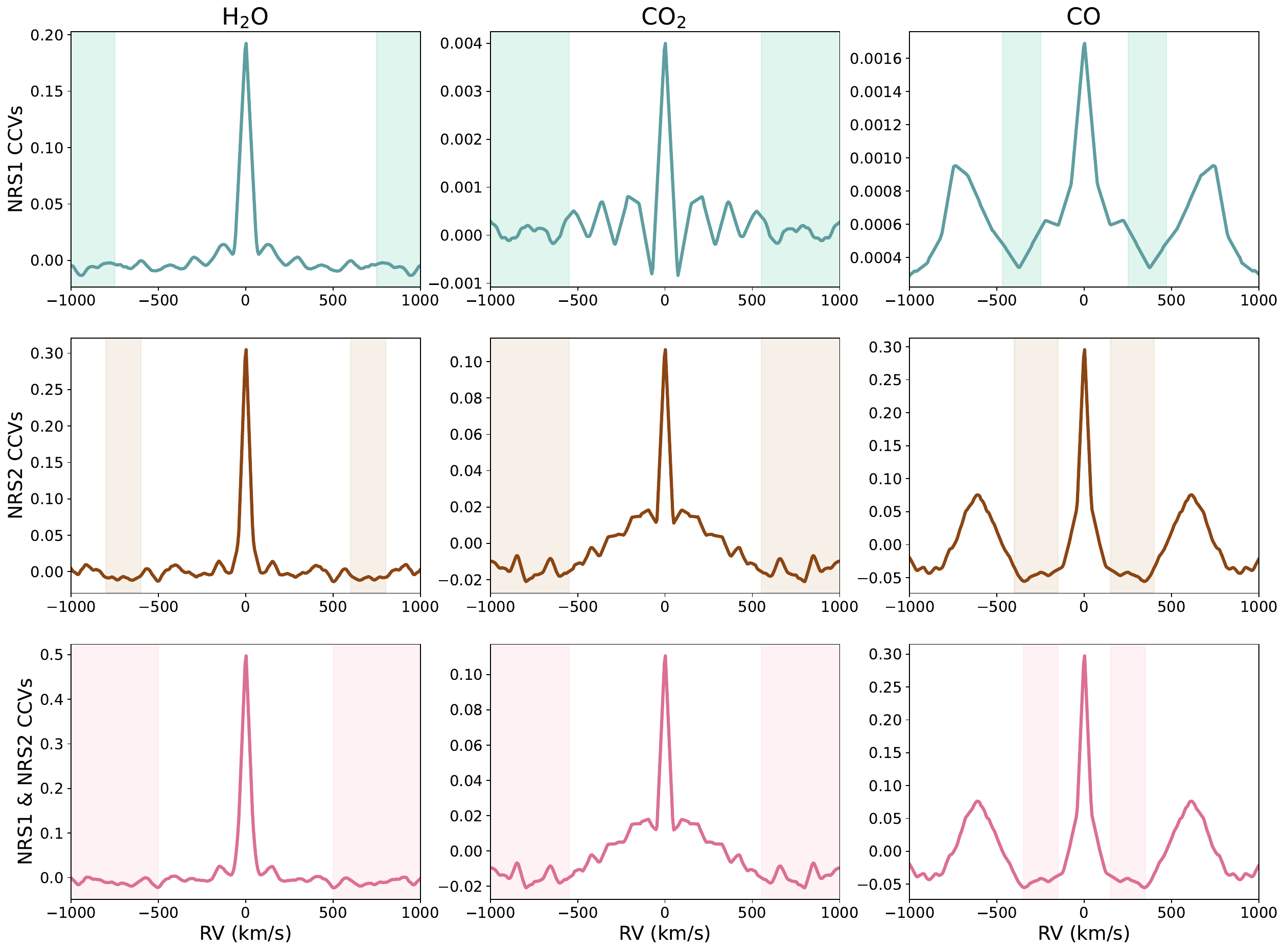}

\caption{Definition of the baseline regions (shaded regions, see Table~\ref{tab:baselines}) with respect to the template-template CCF within the optimal wavelength ranges using the Gaussian normalization on NRS1 (top panels), NRS2 (middle panels) and both ranges combined (bottom panels) for each detected molecule, i.e, $\rm H_{2}O$ (left column), $\rm CO_{2}$ (central column) and $\rm CO$ (right column). In Appendix~\ref{Appendix:baselines_freq} we show an analogous plot considering the frequency filter normalization instead.
}
\label{fig:Baselines}
\end{figure*}

We also performed the cross-correlation of each molecule's template with itself, to identify secondary CCF peaks and to identify regions in velocity space where the cross-correlation values (CCVs) of the CCF are low and can therefore be used as cross-correlation baseline values (see Figure~\ref{fig:Baselines}).

The last step of the method is to define the baseline regions of the CCF for each molecule, with respect to which the significance of the detected peak in the CCF is reported. This is a similar step to the one presented in Sect. 4 of EB23 for the CO band in the NRS2 spectrum, where we defined the baseline of the CCF in 
the low-CCV/valley regions between the central peak and the secondary peaks (if any) in the template-template CCF, which indicate the radial velocity regions (with respect to the velocity of the central peak) where the CCVs of the data-template CCF are expected to be dominated by noise. In Figure~\ref{fig:Baselines}, we show the definition of the baseline regions for the detected molecules ($\rm H_{2}O$, $\rm CO_{2}$ and $\rm CO$) compared with their template-template CCFs.

Additionally, to facilitate the interpretation of the resultant CCFs we converted the CCFs from cross-correlation values (CCVs) to signal-to-noise ratio (SNR) values following the conversion procedure described in Section 4 of EB23 and computing them as:

\begin{equation}
    \rm SNR(RV) = \frac{\mid CCV(RV) - \mu_{baseline} \mid}{\sigma_{baseline}}
\end{equation}
where $\rm \mu_{baseline}$ and $\rm \sigma_{baseline}$ are the mean and standard deviation of the CCVs in the baseline regions.


\section{Cross-correlation results}\label{sec: ccresults}

We carried out cross-correlations between the normalized observed spectrum and the normalized template for each molecule listed in Table~\ref{tab: linelists}. 
For each molecule, we systematically performed cross-correlations across the whole wavelength ranges covered by the NRS1 and NRS2 detectors, first separately (Figures \ref{fig:CCFs_Gauss} and \ref{fig:CCFs_freq}) and then jointly (Figure \ref{fig:CCFs_CombNRS12}). We repeated this process for two different normalization modes: subtracting a Gaussian filter (Figures~\ref{fig:CCFs_Gauss} and \ref{fig:CCFs_CombNRS12}) and filtering by a high-pass frequency filter (Figures~\ref{fig:CCFs_freq} and \ref{fig:CCFs_CombNRS12}). 
For each molecule, we compared the resulting observations-template CCF with the template-template CCF (shown in dashed lines in Figures~\ref{fig:CCFs_H2O_most}, \ref{fig:CCFs_CO2_most}, \ref{fig:CCFs_CO}, \ref{fig:CCFs_Gauss}, \ref{fig:CCFs_freq} and \ref{fig:CCFs_CombNRS12}), with the aim of identifying the presence of secondary peaks in the CCFs. To perform this comparison we put the template-template CCF in the planet's rest frame by introducing an offset in radial velocities that accounts for the systemic velocity of WASP-39b system ($\rm -58.4421~km~s^{-1}$) and the JWST barycentric velocity at the time of the observations ($\rm -28.888~km~s^{-1}$) . 


For each molecule, we performed an additional cross-correlation search over a portion of the observed transmission spectrum that we defined as the optimal wavelength range considering the distribution of lines for a given molecule, and avoiding the wavelength ranges where other molecules might dominate (Figure~\ref{fig:cross-sections_molecules}). The optimal wavelength ranges considered for each molecule are listed in Table~\ref{tab:optimal_wranges}. For instance, we masked the portion of the spectrum between $\rm 4.2~\mu m$ and $\rm 4.6~\mu m$ in the optimal wavelength range definition for all molecules except $\rm CO_{2}$, as in this range the opacity contribution of $\rm CO_{2}$ clearly dominates over any other molecular feature.

\begin{table}
\caption{Definition of the optimal wavelength ranges for each detected molecule.}
\begin{tabular}{lcc}
      \toprule \toprule
      
       & NRS1 Wavelength & NRS2 Wavelength \\ 
      Molecule & Range (${\rm \mu}$m) & Range (${\rm \mu}$m) \\
      \midrule
      
      $\rm H_{2}O$ & $[2.71, 3.72]$ & $[3.82,4.20]\cup[4.60,5.18]$\\
      $\rm CO_{2}$ & $[2.71,3.00]$ & $[4.00,4.80]$\\
      $\rm CO$ & $[2.71,3.20]$ & $[4.60,5.00]$\\
      
      \bottomrule \bottomrule
      
      
          
      \end{tabular}
      
      
      \label{tab:optimal_wranges}
\end{table}

Using our generalized methodology, we confirm the detection of $\rm H_{2}O$, $\rm CO_{2}$ and $\rm CO$ via cross-correlation. Below we discuss the results that achieve the highest detection significance for these molecules in Sections~\ref{sec:H2O},~\ref{sec:CO2} and~\ref{sec:CO}, which are generally obtained from the cross-correlation search over the optimal wavelength range for each molecule. We analyze the significance of these detections in Section~\ref{sec:MCMC}. 
In addition, we show the results of the systematic cross-correlation search throughout the NRS1 and NRS2 detectors in Appendix~\ref{ap: sys_results}.
We report non-detections for the rest of molecular species investigated, which are $\rm CH_{4}$, $\rm NH_{3}$, $\rm SO_{2}$, $\rm N_{2}O$, $\rm H_{2}S$, $\rm PH_{3}$, $\rm O_{3}$ and $\rm C_{2}H_{2}$. These non-detections are in agreement with the expected atmospheric composition of WASP-39b \citep{JWSTERS2023,Rustamkulov2023,Feinstein2023,Alderson2023,Tsai2022,EsparzaBorges23,Grant2023}, except for the non-detection of $\rm SO_{2}$, which is further discussed in Section~\ref{sec:conclusions}.

\subsection{$\rm H_{2}O$ detection}\label{sec:H2O}

Among all studied molecules, water produces the most significant cross-correlation detection. $\rm H_{2}O$ is detected both in the NRS1 ($\rm 2.71\mu m - 3.72\mu m$) and NRS2 ($\rm 3.82\mu m - 5.18\mu m$) detectors independently, as well as in their combined analysis. These cross-correlation functions converted to SNR are shown in the top panels of Figures~\ref{fig:CCFs_Gauss}, \ref{fig:CCFs_freq} and \ref{fig:CCFs_CombNRS12}, respectively, where the SNRs are defined as the ratios between the CCVs and the average CCVs of the baseline regions as defined in Section~\ref{subsec: ccanalysis} and Figure~\ref{fig:Baselines}. These CCFs show central peaks that appear shifted towards the expected radial velocity offset corresponding to the barycentric velocities of WASP-39b system and JWST at the time of the observations.





\begin{figure}
\includegraphics[width=\columnwidth]{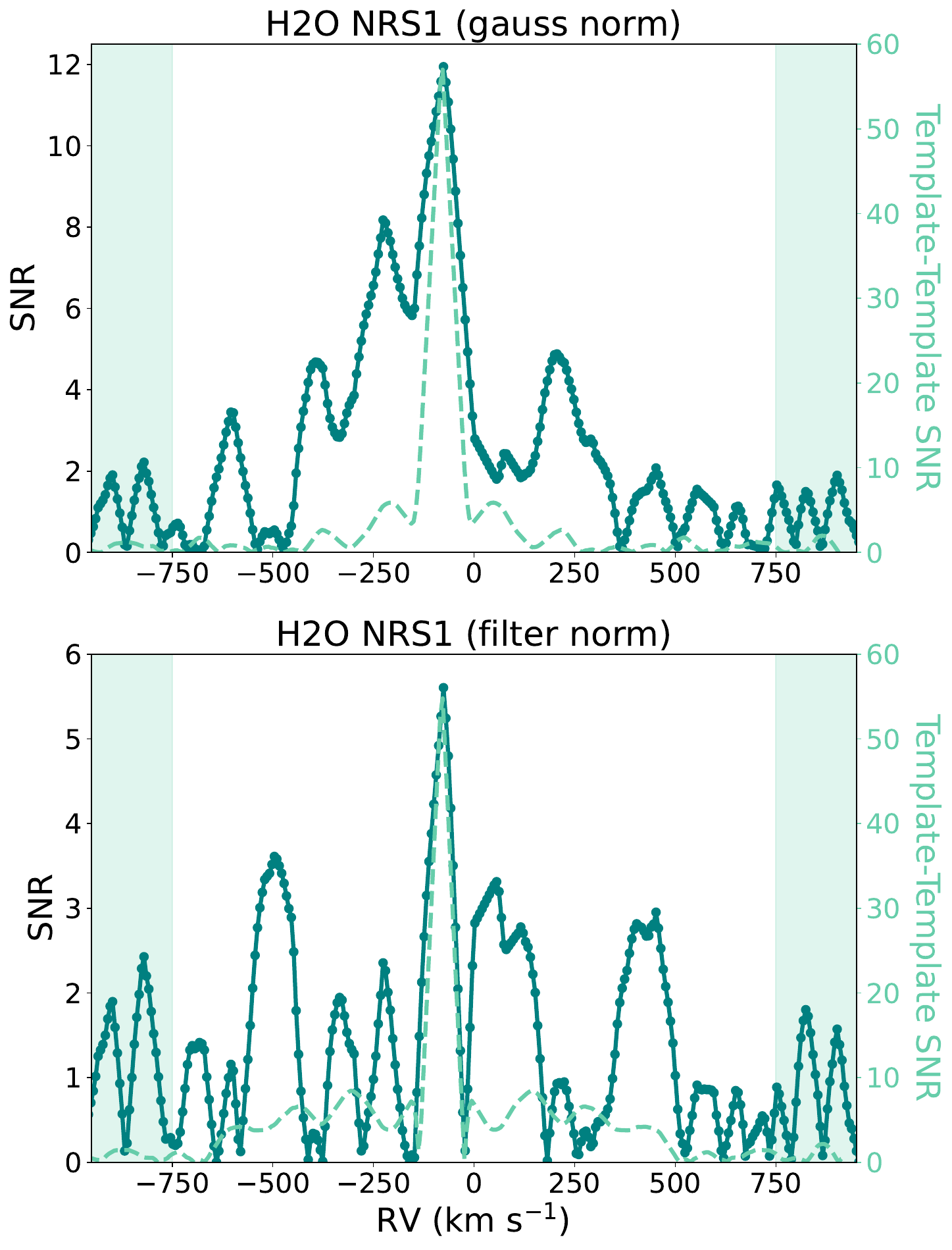}
\caption{Result of the cross-correlation search of $\rm H_{2}O$ along the NRS1 detector ($\rm 2.71 - 3.72~\mu$m) using the Gaussian normalization mode (top) and the frequency-filter normalization mode (bottom). The solid-dotted line shows the resulting data-template CCF and the dashed line shows the template-template CCF shifted towards the expected radial velocity offset produced by the barycentric velocities of WASP-39 system and JWST at the time of the observations.
}
\label{fig:CCFs_H2O_most}
\end{figure}

In Figure~\ref{fig:CCFs_H2O_most} we show the CCF resulting from the search throughout the NRS1 detector's wavelength range compared to the shifted template-template CCF for the same wavelength range, where the $\rm H_{2}O$ detection is the most significant for both normalization modes. We obtained that the central peak of the CCF is more prominent when using the Gaussian fitting subtraction normalization ($\rm SNR=11.9$) than with the frequency filtering normalization ($\rm SNR=5.6$). Besides this, we find that $\rm H_{2}O$ is confidently detected in the NRS1 detector regardless of the normalization mode while in the NRS2 the detection is strongly favoured by the use of the Gaussian normalization mode.



\subsection{$\rm CO_{2}$ detection}\label{sec:CO2}

The presence of $\rm CO_{2}$ is detected in the NRS2 detector and in the combination of the NRS1 and NRS2 detectors with both normalization modes. However, the signal of $\rm CO_{2}$ is not retrieved by cross-correlations in the NRS1 detector's wavelength range alone through any of the normalization modes. This is expected because of the predominance of $\rm H_{2}O$ lines in the NRS1 spectral range, which are probably diluting the cross-correlation signal of $\rm CO_{2}$, and also because the blue edge of the NRS1 detector is a very noisy region where the observed spectrum has a low SNR. 



\begin{figure}
\includegraphics[width=\columnwidth]{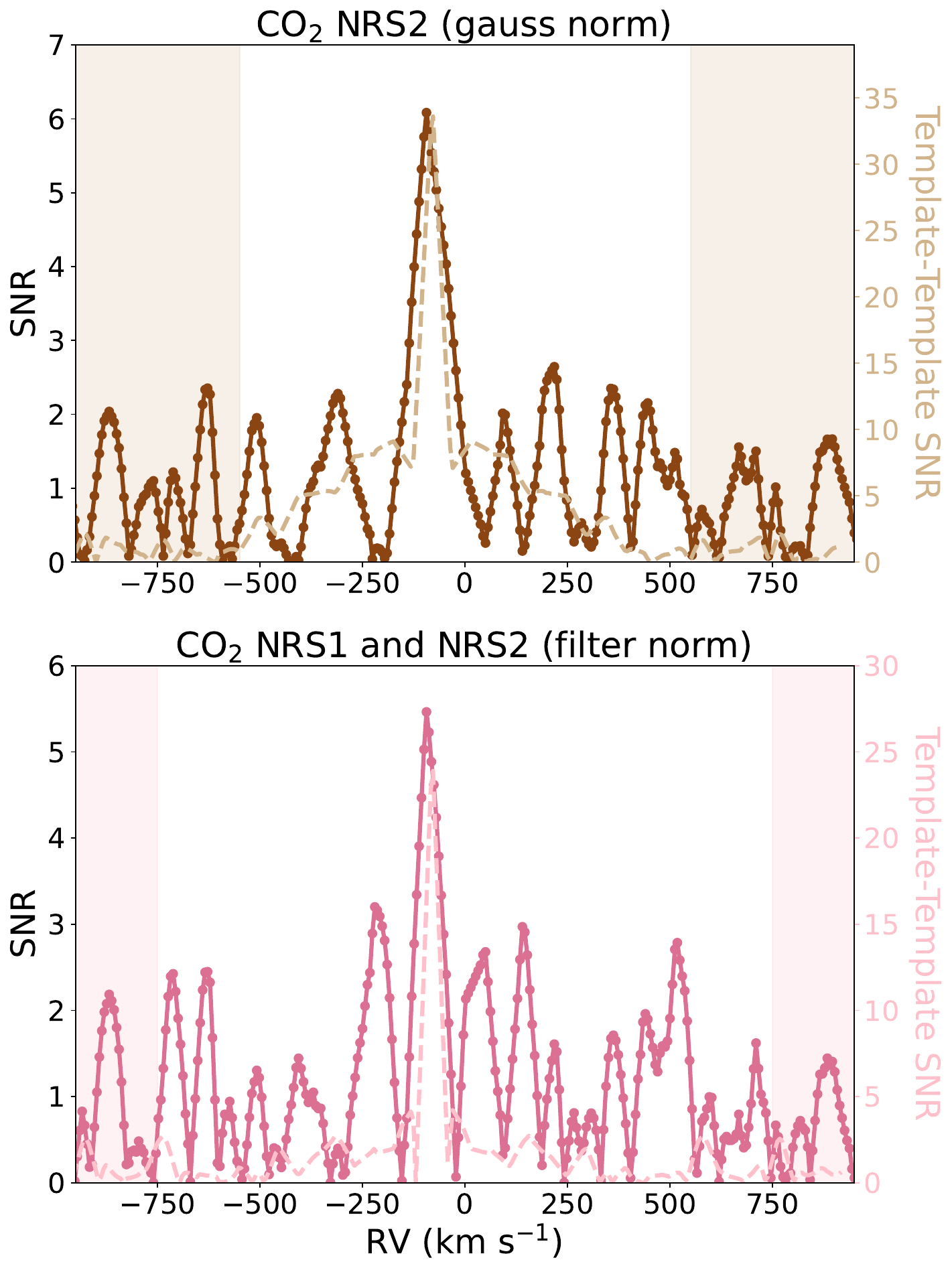}
\caption{Results of the cross-correlation search of $\rm CO_{2}$ along the optimal wavelength range of NRS2 detector using the Gaussian normalization mode (top) and along the wavelength ranges of NRS1 and NRS2 detectors using frequency-filter normalization (bottom). The solid-dotted line shows the data-template CCF and the dashed line shows the template-template CCF.
}
\label{fig:CCFs_CO2_most}
\end{figure}

In Figure~\ref{fig:CCFs_CO2_most} we show the cross-correlation result on NRS2 detector using the Gaussian normalization mode (top panel) and on the combination of the NRS1 and NRS2 detectors with the frequency filter normalization (bottom panel), which correspond to the most significant $\rm CO_{2}$ detections for each normalization mode with $\rm SNR=6.1$ and $\rm SNR=5.5$, respectively. Both results are obtained using the optimal wavelength range definition (Table~\ref{tab:optimal_wranges}), where the $\rm CO_{2}$ opacity contribution is expected to be significant or dominant considering the rest of the expected molecules.


\subsection{$\rm CO$ detection}\label{sec:CO}

\begin{figure}
\includegraphics[width=\columnwidth]{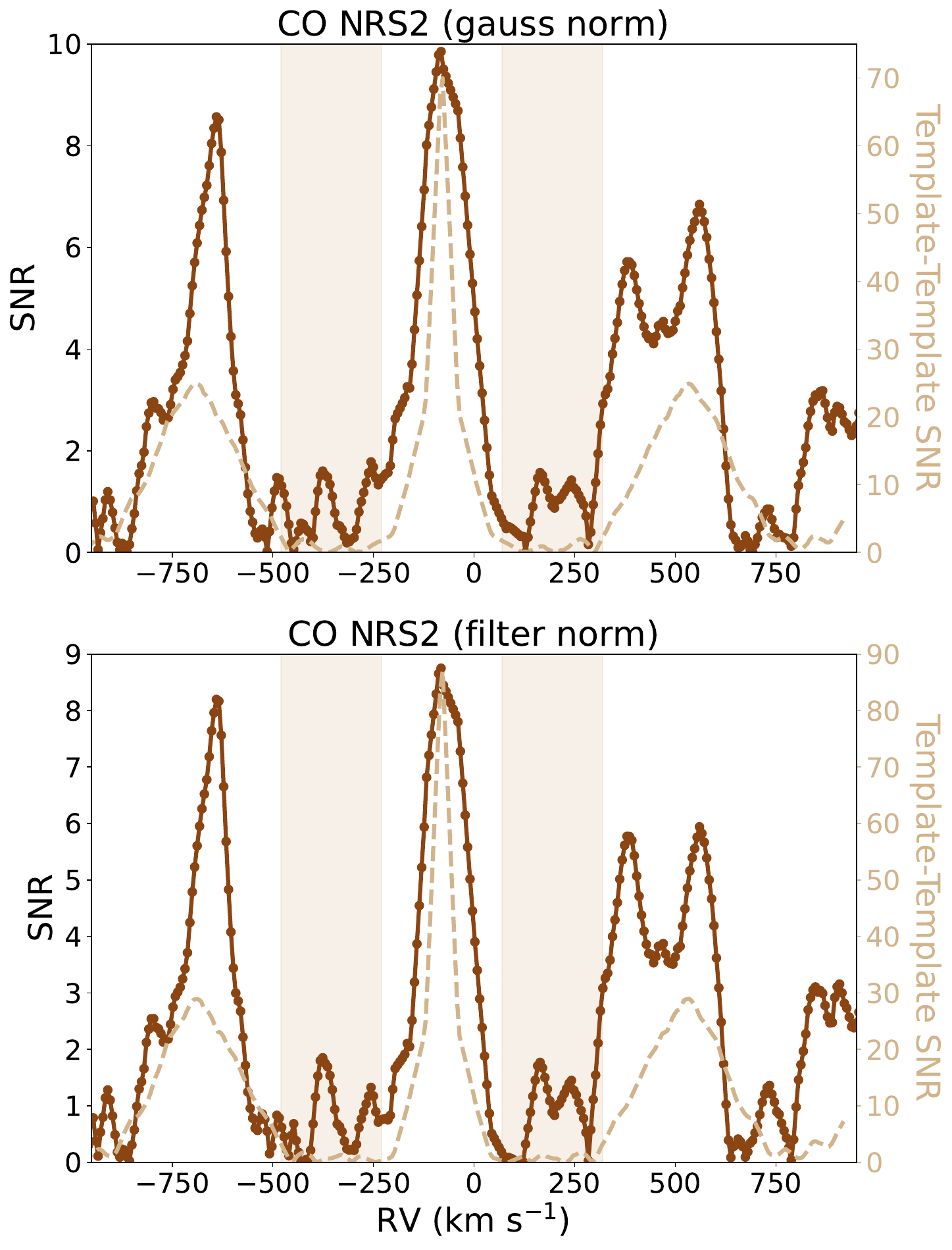}
\caption{Results of the cross-correlation search of $\rm CO$ along the optimal wavelength range for NRS2 detector ($\rm 4.6 - 5.0 \mu m$) using the Gaussian normalization mode (top) and the frequency-filter normalization mode (bottom). The solid-dotted line shows the data-template CCF and the dashed line shows the template-template CCF.
}
\label{fig:CCFs_CO}
\end{figure}

We confirm the previous detections of CO \citep{Grant2023,EsparzaBorges23} through this improved cross-correlation methodology, achieving a similar detection confidence than that obtained in EB23. CO is detected in NRS2 detector and in the combination of NRS1 and NRS2 detectors with both normalization modes. Figure~\ref{fig:CCFs_CO} shows the cross-correlation result on the optimal wavelength range of NRS2 detector, ($\rm 4.6 - 5.0 \mu m$), using the Gaussian normalization (top panel) and the frequency filter normalization (bottom panel). In this work, we generated only one CO template using a linelist containing all CO isotopologues, as here we focus on testing the performance of the technique among different molecules. The selection of different CO isotopologues for the generation of different CO templates and their impact on the cross-correlation results were preliminarily analysed in EB23 and will be further studied in future dedicated articles.



\subsection{MCMC sampling}\label{sec:MCMC}


The relative SNR of the CCF peaks shown in Figures \ref{fig:CCFs_H2O_most}, \ref{fig:CCFs_CO2_most} and \ref{fig:CCFs_CO} depends on the 'instantaneous' noise structure of the transmission spectrum as observed. Therefore, to evaluate the significance of the reported SNRs for the detected molecules we performed a Monte-Carlo Markov Chain (MCMC) procedure, where each point on the transmission spectrum is resampled randomly within the transit depth uncertainties, generating 2000 samples of the transmission spectrum. Then, we performed the cross-correlation of each sampled transmission spectrum with each molecular template and converted the resulting CCF to SNR following the same methodology described in Section~\ref{sec: methodology}.



\begin{figure*}
\includegraphics[width=\textwidth]{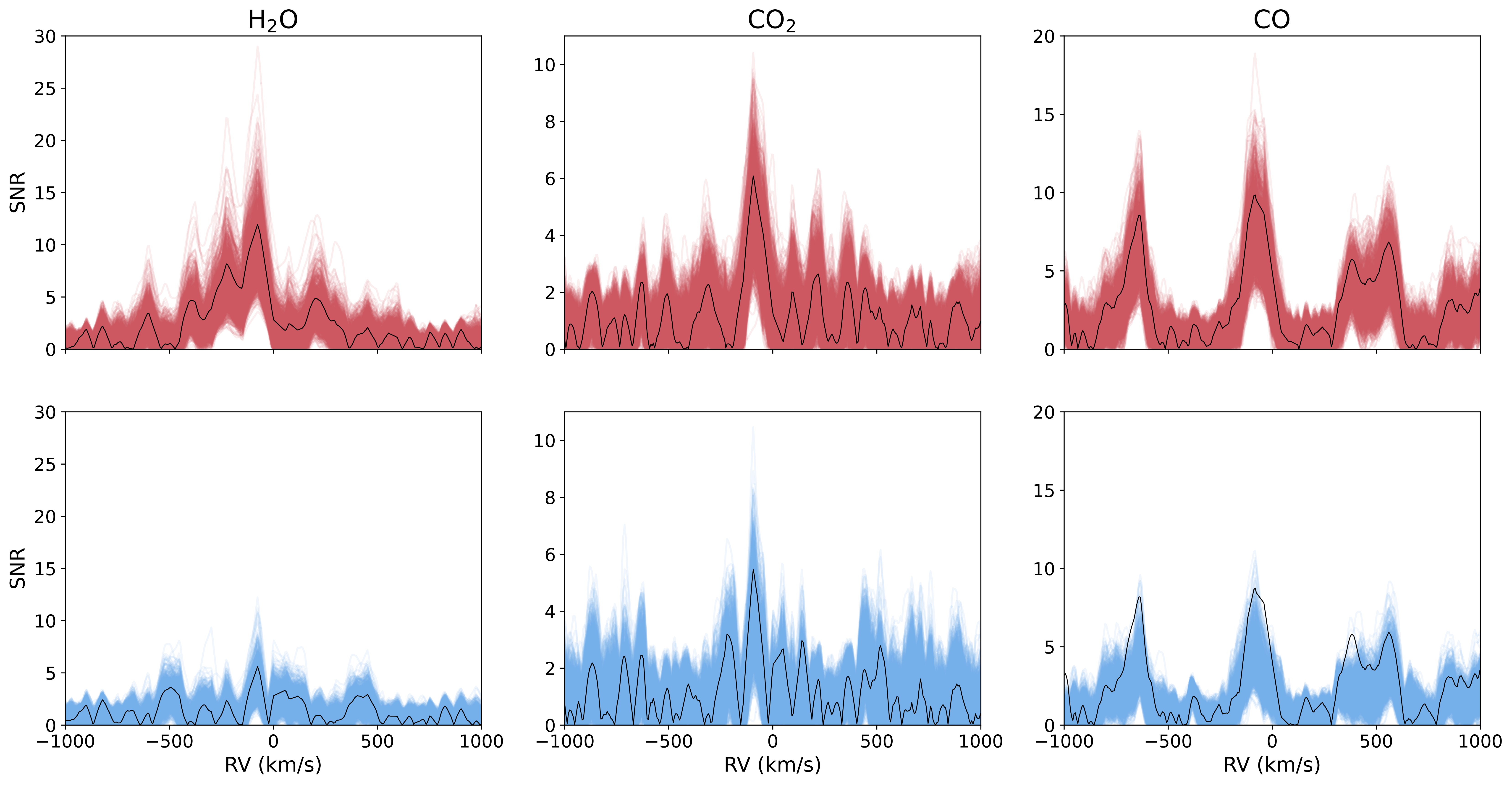}
\caption{Results of the cross-correlation search of $\rm H_{2}O$ (left panels), $\rm CO_{2}$ (middle panels) and $\rm CO$ (right panels) along their optimal wavelength ranges over the 2000 samples of the transmission spectrum (those shown in Figure~\ref{fig:CCFs_H2O_most}, Figure~\ref{fig:CCFs_CO2_most} and Figure~\ref{fig:CCFs_CO}). Black-solid lines show the CCFs resulting from the cross-correlation search over the original transmission spectrum. Top panels show the results using the Gaussian normalization mode and bottom panels show the results using the frequency filter normalization mode.
}
\label{fig:CCFs_ALLBEST}
\end{figure*}

In Figure~\ref{fig:CCFs_ALLBEST}, we show the results of the cross-correlations search of $\rm H_{2}O$, $\rm CO_{2}$ and $\rm CO$ (along their optimal wavelength ranges) for the 2000 transmission spectrum samples, using the Gaussian fitting normalization (top panels) and the frequency filtering normalization (bottom panels). 

\begin{figure*}
\includegraphics[width=\textwidth]{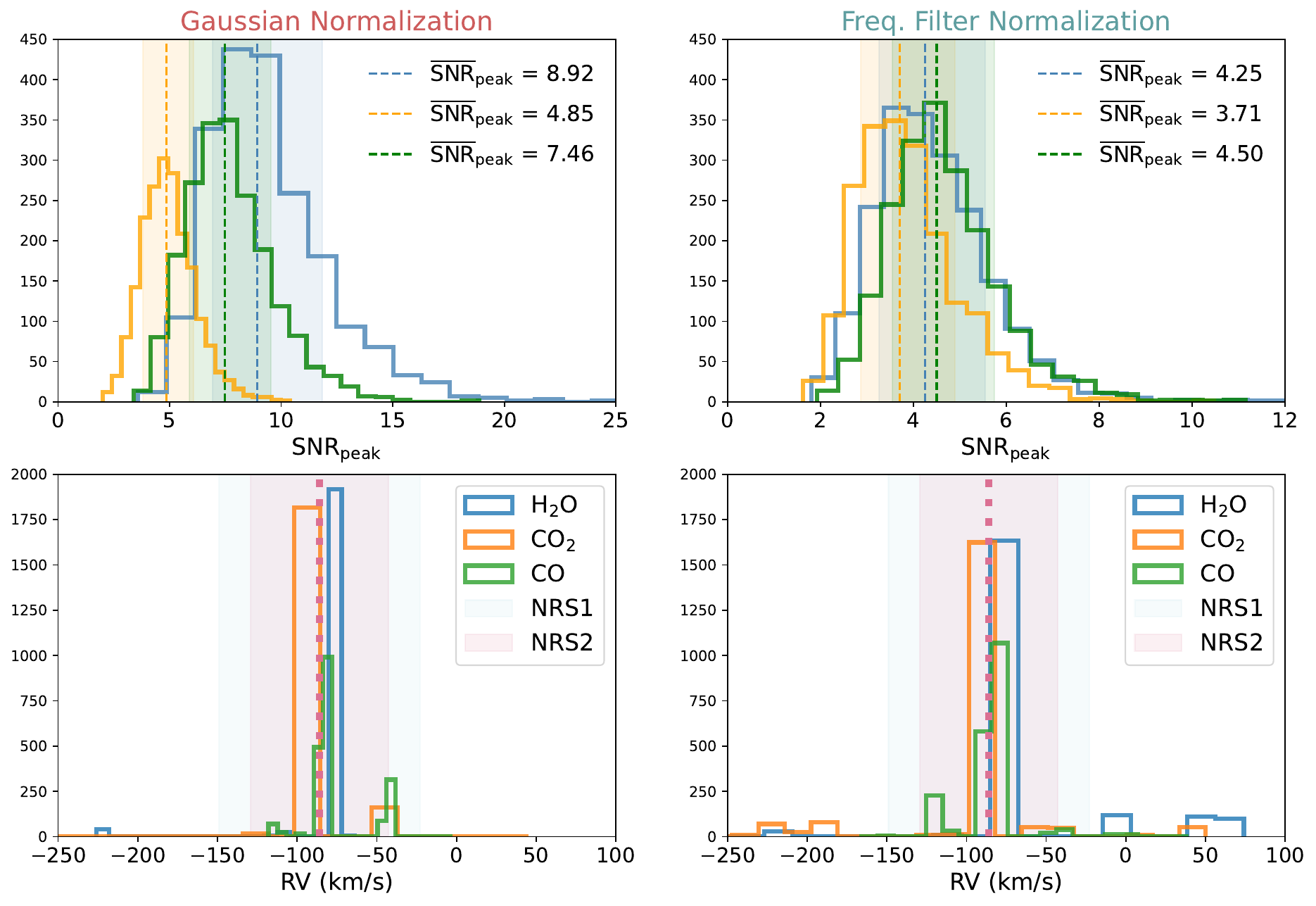}
\caption{Distribution of the CCF central peak maximum SNR values ($\rm SNR_{peak}$) and their RV positions from the cross-correlation results over the transmission spectrum samples for each detected molecule and for both normalization modes. The top panels show the $\rm SNR_{peak}$ distribution for each detected molecule, indicating their median values (dashed line) and their $1\sigma$ percentiles (shaded regions), and the bottom panels show the distribution of the RV position of the central peak's maximum value for each molecule compared to the expected systemic velocity of WASP-39b relative to JWST at the time of the observations ($\rm -87~km/s$, pink dashed line) and the mean RV pixel size of each detector (shaded regions). The left panels show the results for the Gaussian normalization mode and the right panels show the results for the frequency filter normalization mode.
}
\label{fig:Histograms_peaks}
\end{figure*}

For each sampled CCF we identified the maximum SNR value of the central peak ($\rm SNR_{peak}$) and its RV position. Figure~\ref{fig:Histograms_peaks} shows the sampled distribution of $\rm SNR_{peak}$ values and the distribution of RV positions at which the maxima of the central peak appear, for each detected molecule and for both normalization modes. To determine the statistical significance of each detection, we measured the median value and $\rm 1\sigma$ percentiles on each $\rm SNR_{peak}$ distribution. In Table~\ref{tab:significance_results} we report the significance of each molecular detection by indicating the median $\rm SNR_{peak}$ and $\rm 1\sigma$ percentiles for each molecule. These results show that the Gaussian fitting achieves the highest detection significance for all molecules, being the most efficient normalization mode. 



\begin{table}
\caption{Median of $\rm SNR_{peak}$ for each detected molecule and their $\rm 1\sigma$ uncertainties. The columns show this result for both normalization modes (Gaussian and frequency filter).}
\begin{tabular}{lcc}
      \toprule \toprule
      
       & $\rm SNR_{peak}$ & $\rm SNR_{peak}$ \\ 
      Molecule & (Gaussian) & (Freq. filter) \\
      \midrule
      
      $\rm H_{2}O$ & $\rm 8.9^{+2.9}_{-2.0}$ & $\rm 4.3^{+1.3}_{-1.0}$\\
      $\rm CO_{2}$ & $\rm 4.9^{+1.2}_{-1.0}$ & $\rm 3.7^{+1.2}_{-0.8}$\\
      $\rm CO$ & $\rm 7.5^{+2.1}_{-1.6}$ & $\rm 4.5^{+1.2}_{-1.0}$\\
      
      \bottomrule \bottomrule
      
      
          
      \end{tabular}
      
      
      \label{tab:significance_results}
\end{table}


\section{Conclusions and Discussion}\label{sec:conclusions}

We present an improved, generalized cross-correlation methodology for the search of molecular species in 
JWST transmission spectra. We applied this methodology to the WASP-39 b JTEC ERS Program NIRSpec G395H dataset. 

We demonstrate that cross-correlation effectively detects multiple molecular species in JWST transmission spectra, proving particularly effective for molecules with mid-to-low SNR and narrow spectral features (e.g., $\rm H_{2}O$, $\rm CO$). This technique is more sensitive to individual spectral lines than to broad molecular features. The performance of this technique is limited for the detection of broad spectral features because the normalization step removes low-frequency signals characteristic of a specific molecule, e.g., Gaussian-like features in $\rm CO_{2}$ around $\rm 4.4~\mu m$ or $\rm SO_{2}$ around $\rm 4.0~\mu m$ and slopes in $\rm H_{2}O$. However, this fact is also an advantage, since the remaining high-frequency pattern signals of those absorptions are less prone to systematics (e.g., bumps in the spectrum).

We confirm the detections of $\rm H_{2}O$, $\rm CO_{2}$ and $\rm CO$, achieving detection significances of $\rm SNR_{peak}=8.9 ^{+2.9}_{-2.0}$, $\rm SNR_{peak}=4.9 ^{+1.2}_{-1.0}$ and $\rm SNR_{peak}=7.5 ^{+2.1}_{-1.6}$, respectively. Our analysis shows that -- for all detected molecules -- the use of the Gaussian normalization mode for the generation of templates leads to higher detection significances than the frequency filter normalization mode.

We find that the detection significance does not have any substantial variation with the selection of the cross-correlation RV step. However, we note a slight increase in the uncertainties in Table~\ref{tab: RV_steps}, especially in the case of $\rm H_{2}O$ for the 64~km/s step, which might be related to an aliasing effect that depends on the density of the spectral lines. This effect needs to be explored in more detail in future studies, in particular studying how it might affect some molecules more than others and how it impacts the SNR values estimated using the CCF peak divided by the standard deviation of a predefined baseline.

Our results show that the molecular detections are more efficient when doing the cross-correlation search over the optimal wavelength ranges, which are defined as the spectral regions where the contribution to the opacity of each molecular feature is expected to be dominant or considerable considering the relative contribution of the rest of molecules present in the spectrum.


Exoplanet atmospheres are composed by multiple molecular species, and the signals of multiple molecules can overlap, specially at the lower spectral resolution levels produced by JWST observations. This introduces additional noise in the CCFs that can lower the significance of the detections. However, this is not an issue in terms of identifying the individual absorptions pattern of an specific molecule. 
While this study focuses on identifying individual molecules in a transmission spectrum through cross-correlations, the developed methodology can be expanded to the search for combinations of molecules, which would allow measuring (or placing limits) in abundance ratios.


\section*{Acknowledgements}

This work is based on observations made with the NASA/ESA/CSA JWST. The data were obtained from the Mikulski Archive for Space Telescopes at the Space Telescope Science Institute, which is operated by the Association of Universities for Research in Astronomy, Inc., under NASA contract NAS 5-03127 for JWST. These observations are associated with program JWST-ERS-01366. Support for program JWST-ERS-01366 was provided by NASA through a grant from the Space Telescope Science Institute. This work was supported by grant JWST-ERS-01366.033-A. E.E-B. and E.P. acknowledge funding from the Spanish Ministry of Economics and Competitiveness through project PGC2018-098153-B-C31. E.E-B. acknowledges financial support from the European Union and the State Agency of Investigation of the Spanish Ministry of Science and Innovation (MICINN) under the grant PRE2020-093107 of the Pre-Doc Program for the Training of Doctors (FPI-SO) through FSE funds. J.K. acknowledges financial support from Imperial College London through an Imperial College Research Fellowship grant. C.C. acknowledges support by ANID BASAL project FB210003. L.D. acknowledges support from the KU Leuven IDN grant IDN/19/028 and the MC-ITN CHAMELEON grant. This research was supported by the Excellence Cluster ORIGINS which is funded by the Deutsche Forschungsgemeinschaft (DFG, German Research Foundation) under Germany's Excellence Strategy - EXC-2094 - 390783311. G.M. acknowledges financial support from the Severo Ochoa grant CEX2021-001131-S and from the Ramón y Cajal grant RYC2022-037854-I funded by MCIN/AEI/1144 10.13039/501100011033 and FSE+. This material is based upon work supported by the National Aeronautics and Space Administration under Agreement No.\ 80NSSC21K0593 for the program ``Alien Earths''. The results reported herein benefited from collaborations and/or information exchange within NASA’s Nexus for Exoplanet System Science (NExSS) research coordination network sponsored by NASA’s Science Mission Directorate. J-M.D. acknowledges the research program VIDI New Frontiers in Exoplanetary Climatology with project number 614.001.601, which is (partly) financed by the Dutch Research Council (NWO).



\section*{Data Availability}
The data were obtained from the Mikulski Archive for Space Telescopes at the Space Telescope Science Institute, which is operated by the Association of Universities for Research in Astronomy, Inc., under NASA contract NAS 5-03127 for JWST. The specific observations can be accessed via \url{http://dx.doi.org/10.17909/gdm2-0q65}. These observations are associated with program JWST-ERS-01366.
 



\bibliographystyle{mnras}
\bibliography{example} 




\appendix

\section{Impact of RV sampling in the SNR of the detections}\label{ap: rvsteps}

We repeated the analysis of the detected molecules varying the RV step in the cross-correlations, testing 6, 20, 43 and 64 km/s. We found that the detection significance does not have any substantial variation with the selection of the cross-correlation RV step. We note a slight increase in the uncertainties in Table~\ref{tab: RV_steps}, especially in the case of $\rm H_{2}O$ for the 64~km/s step, which might be related to an aliasing effect that depends on the density of the spectral lines. This effect needs to be explored in more detail in future studies, in particular studying how it might affect some molecules more than others and how it impacts the SNR values estimated using the CCF peak divided by the standard deviation of a predefined baseline.

\begin{figure*}

    \includegraphics[width=\textwidth]{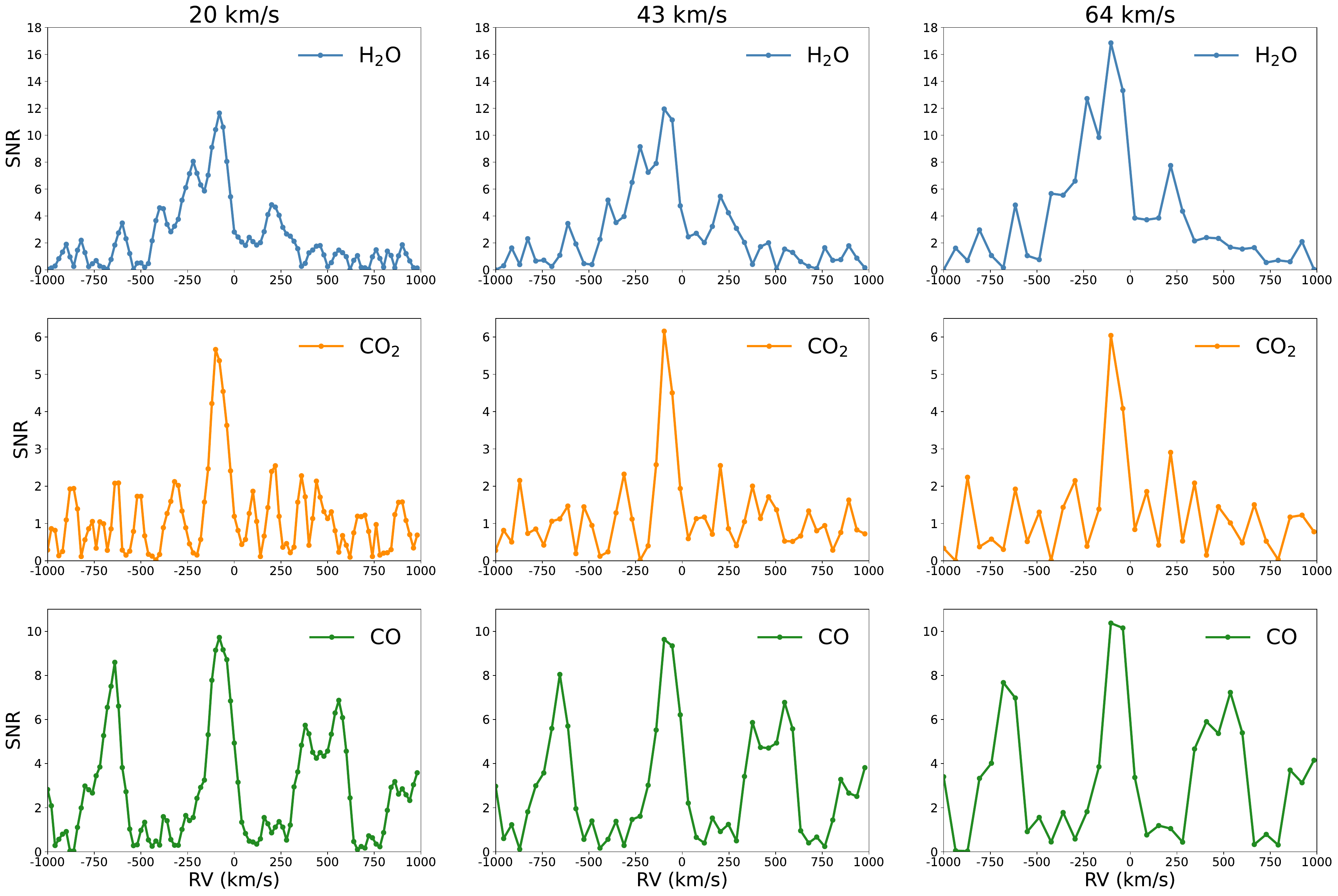}
    \caption{ Signal-to-noise ratio (SNR) of the cross-correlation function (CCF) as a function of radial velocity (RV) for H$_2$O (top row), CO$_2$ (middle row), and CO (bottom row) across three different CCF sampling intervals: 20 km/s (left), 43 km/s (middle), and 64 km/s (right). All panels use the Gaussian normalization mode and apply the cross-correlation over the optimal wavelength ranges defined for each molecule. The consistent detection peaks at the expected velocities across different RV steps demonstrate the robustness of the molecular detections and show that SNR estimates are not significantly affected by the choice of velocity sampling.}
    \label{fig: app_RVsteps_CCFs}
\end{figure*}

\begin{figure*}
    \includegraphics[width=\columnwidth]{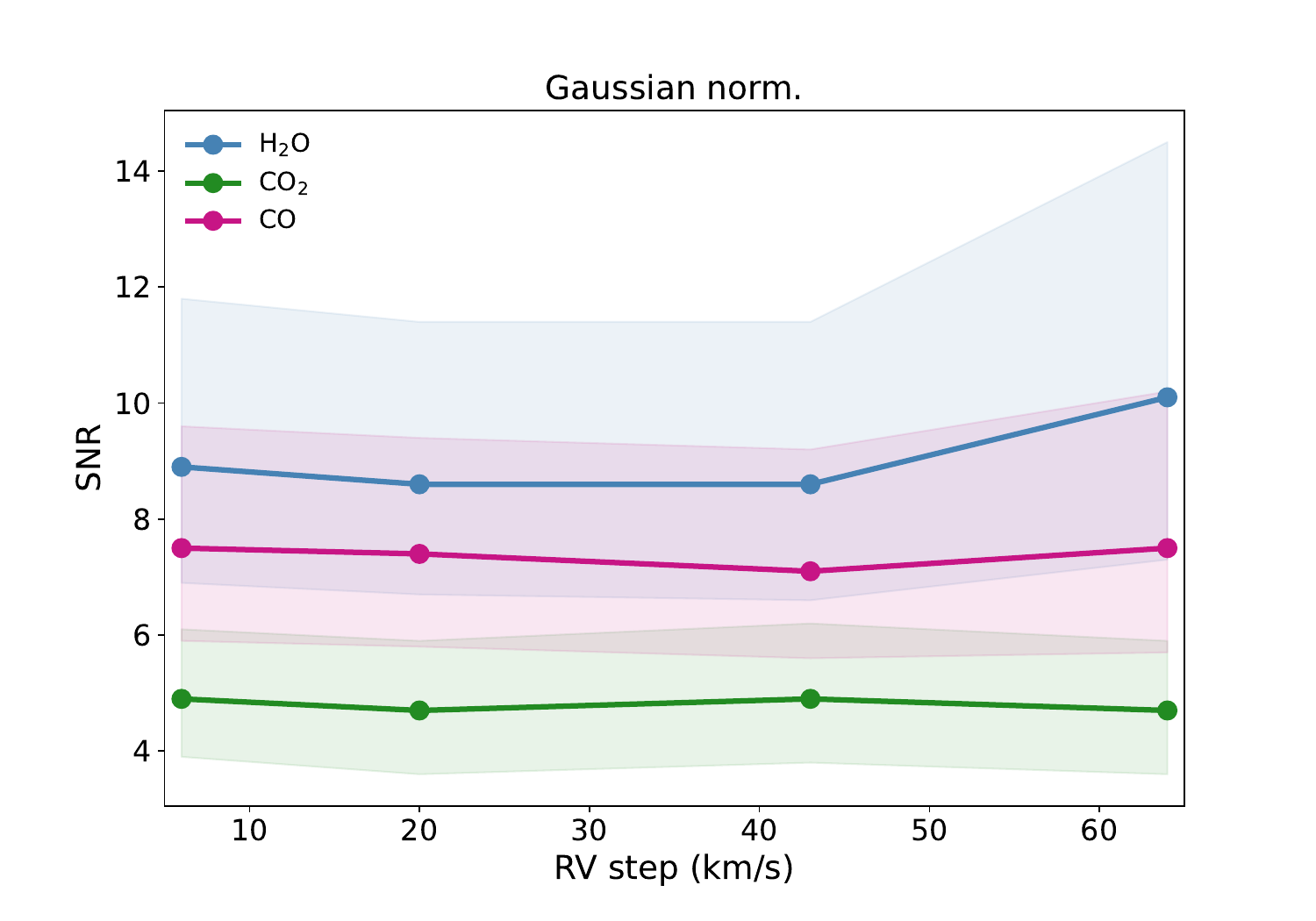}
    \includegraphics[width=\columnwidth]{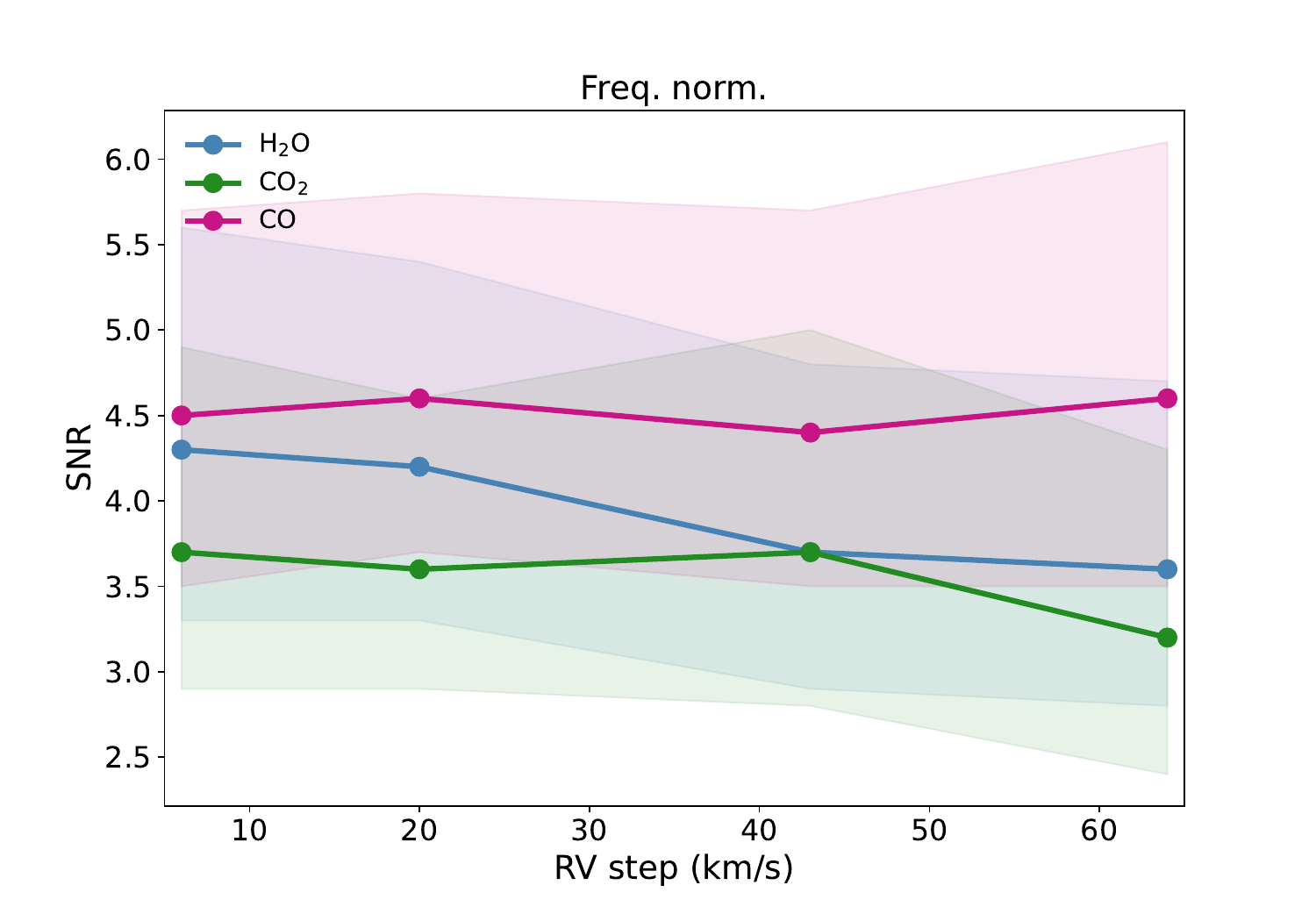}

    \caption{Median SNR values of the CCF central peaks for H$_2$O (blue), CO$_2$ (green), and CO (magenta) as a function of RV sampling step size (6–64 km/s), using Gaussian normalization (left) and frequency-filter normalization (right). Shaded regions indicate the $1\sigma$ intervals from 2000 Monte Carlo samples of the transmission spectrum. The consistency of the SNR values across different RV steps confirms that the detection significance is not strongly affected by the choice of velocity resolution, validating the robustness of the molecular detections.}
    \label{fig: app_RVsteps}
    
\end{figure*}

\begin{table*}\centering
\caption{Peak signal-to-noise (SNR$_\mathrm{peak}$) values for the cross-correlation detection of H$_2$O, CO$_2$, and CO across different RV step sizes (6–64 km/s), using Gaussian (Gauss) and frequency-filter (Freq) normalization modes.}\label{tab: RV_steps}
\begin{tabular}{lrrrrrrrrr}\toprule
&\multicolumn{8}{c}{$\rm SNR_{peak}$} \\\midrule
Molecule &64 km/s (Gauss) &64 km/s (Freq) &43 km/s (Gauss) &43 km/s (Freq) &20 km/s (Gauss) &20 km/s (Freq) &6 km/s (Gauss) &6 km/s (Freq) \\
$\rm H_{2}O$ &$\rm 10.1 ^{+4.4}_{-2.8}$ &$\rm 3.6 ^{+1.1}_{-0.8}$ &$\rm 8.6 ^{+2.8}_{-2.0}$ &$\rm 3.7 ^{+1.1}_{-0.8}$ &$\rm 8.6 ^{+2.8}_{-1.9}$&$\rm 4.2 ^{+1.2}_{-0.9}$ &$\rm 8.9 ^{+2.9}_{-2.0}$ &$\rm 4.3 ^{+1.3}_{-1.0}$ \\[3pt]
$\rm CO_{2}$ &$\rm 4.7 ^{+1.2}_{-1.1}$ &$\rm 3.2 ^{+1.1}_{-0.8}$ &$\rm 4.9 ^{+1.3}_{-1.1}$ &$\rm 3.7 ^{+1.3}_{-0.9}$ &$\rm 4.7 ^{+1.2}_{-1.1}$ &$\rm 3.6 ^{+1.0}_{-0.7}$ &$\rm 4.9 ^{+1.2}_{-1.0}$ &$\rm 3.7 ^{+1.2}_{-0.8}$ \\[3pt]
CO &$\rm 7.5 ^{+2.7}_{-1.8}$ &$\rm 4.6 ^{+1.5}_{-1.1}$ &$\rm 7.1 ^{+2.1}_{-1.5}$ &$\rm 4.4 ^{+1.3}_{-0.9}$ &$\rm 7.4 ^{+2.0}_{-1.6}$ &$\rm 4.6 ^{+1.2}_{-0.9}$ &$\rm 7.5 ^{+2.1}_{-1.6}$ &$\rm 4.5 ^{+1.2}_{-1.0}$ \\
\bottomrule
\end{tabular}
\end{table*}

\section{Cross-section molecular contributions}

\begin{figure*}
\includegraphics[width=\textwidth]{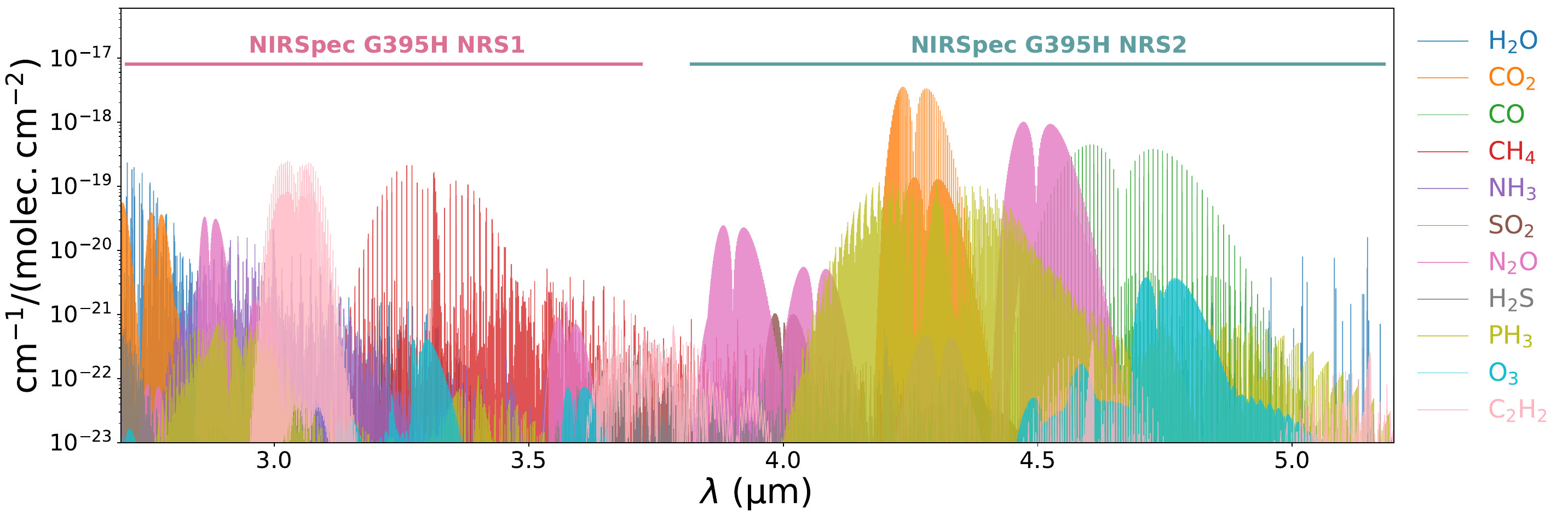}
\caption{Line-intensities of all the molecules considered in this study. The cross-sections are obtained from the HITEMP and HITRAN databases as listed in Table 1.}
\label{fig:cross-sections_molecules}
\end{figure*}

\section{Template-template CCFs and baseline definitions for the frequency filter normalization}\label{Appendix:baselines_freq}

\begin{figure*}
\includegraphics[width=\textwidth]{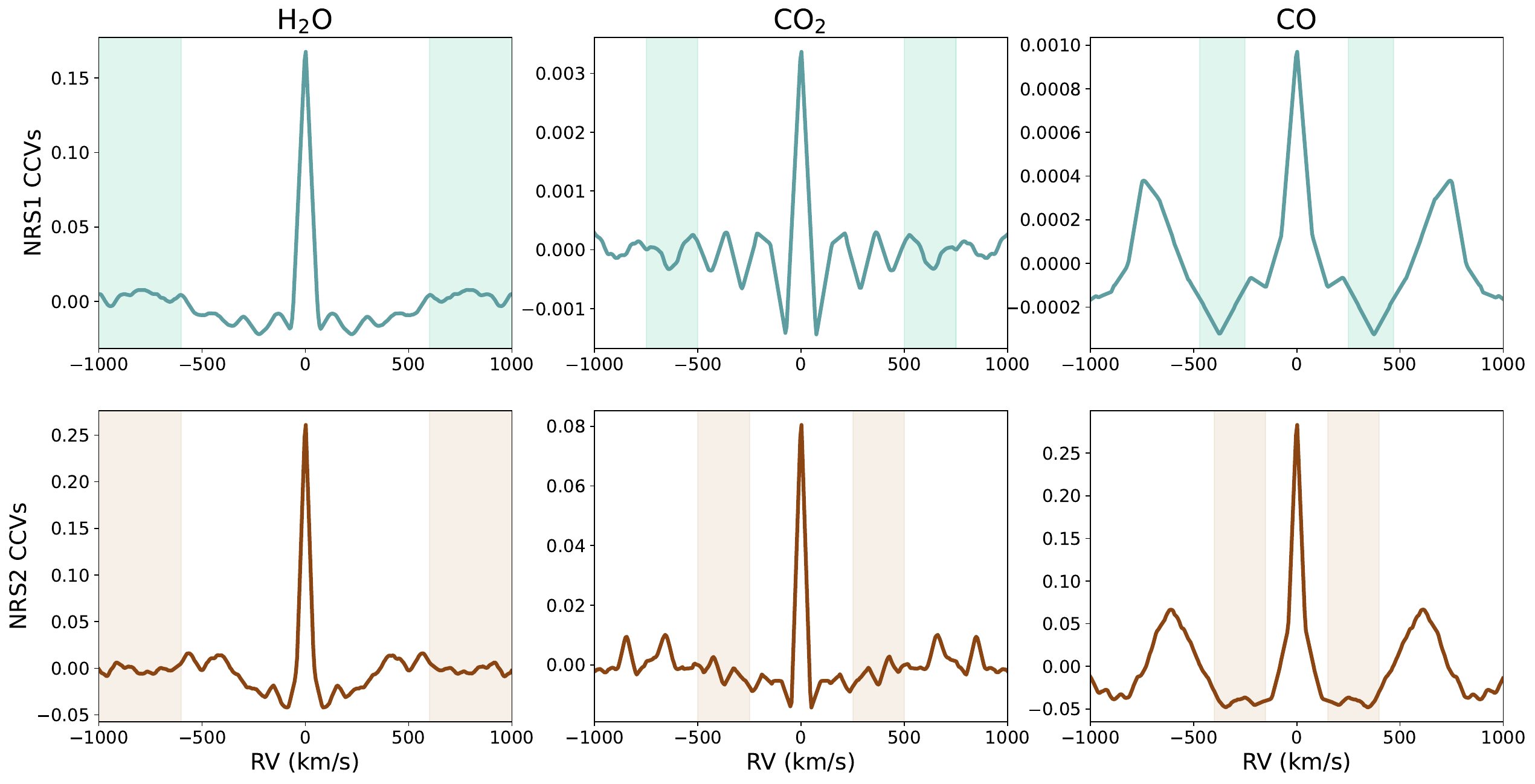}

\caption{Definition of the baseline regions (shaded regions) with respect to the template-template CCF within the optimal wavelength ranges using the frequency filter normalization on the NRS1 (top panels) and the NRS2 (bottom panels) for each detected molecule.
}
\label{fig:Freq_Baselines}
\end{figure*}

\section{Template-template CCFs and baseline definitions for the non-detected molecules}\label{Appendix:baselines_nondected}

\begin{figure*}
\includegraphics[width=\textwidth]{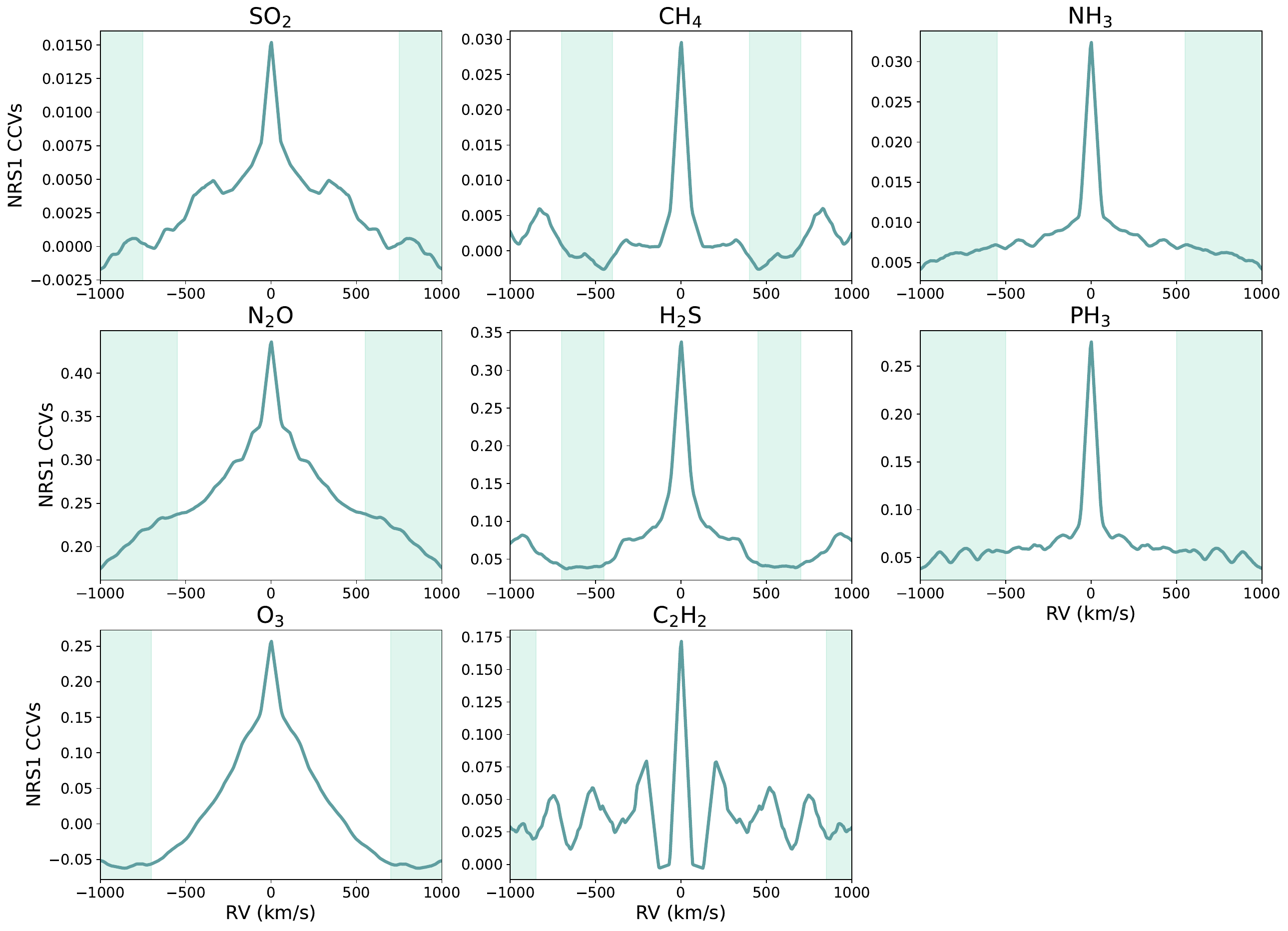}

\caption{Definition of the baseline regions (shaded regions) with respect to the template-template CCF within the optimal wavelength ranges using the Gaussian normalization on NRS1 for the non-detected molecules.
}
\label{fig:Freq_Baselines_nondetected_NRS1}
\end{figure*}

\begin{table}
\caption{Definition of the RV baselines for non-detected molecules.}
\begin{tabular}{lcc}
      \toprule \toprule
      
       Molecule & NRS1 RV (km/s) & NRS2 RV (km/s) \\ 
      \midrule

      $\rm H_{2}O$ & $[-1000,-750] \cup [750,1000]$ & $[-800,-600] \cup [600,800]$\\
      $\rm CO_{2}$ & $[-1000,-550] \cup [550,1000]$ & $[-1000,-550] \cup [550,1000]$\\
      $\rm CO$ & $[-470,-250] \cup [250,470]$ & $[-400,-150] \cup [150,400]$\\
      $\rm SO_{2}$ & $[-1000,-750] \cup [750,1000]$ & $[-1000,-750] \cup [750,1000]$\\
      $\rm CH_{4}$ & $[-700,-400] \cup [400,700]$ & $[-1000,-750] \cup [750,1000]$\\
      $\rm NH_{3}$ & $[-1000,-550] \cup [550,1000]$ & $[-1000,-550] \cup [550,1000]$\\
      $\rm N_{2}O$ & $[-1000,-550] \cup [550,1000]$ & $[-1000,-750] \cup [750,1000]$\\
      $\rm H_{2}S$ & $[-700,-450] \cup [450,700]$ & $[-1000,-750] \cup [750,1000]$\\
      $\rm PH_{3}$ & $[-1000,-500] \cup [500,1000]$ & $[-700,-300] \cup [300,700]$\\
      $\rm O_{3}$ & $[-1000,-700] \cup [700,1000]$ & $[-1000,-700] \cup [700,1000]$\\
      $\rm C_{2}H_{2}$ & $[-1000,-850] \cup [850,1000]$ & $[-1000,-800] \cup [800,1000]$\\
      
      \bottomrule \bottomrule
      
      
          
      \end{tabular}
      
      
      \label{tab:baselines}
\end{table}

\begin{figure*}
\includegraphics[width=\textwidth]{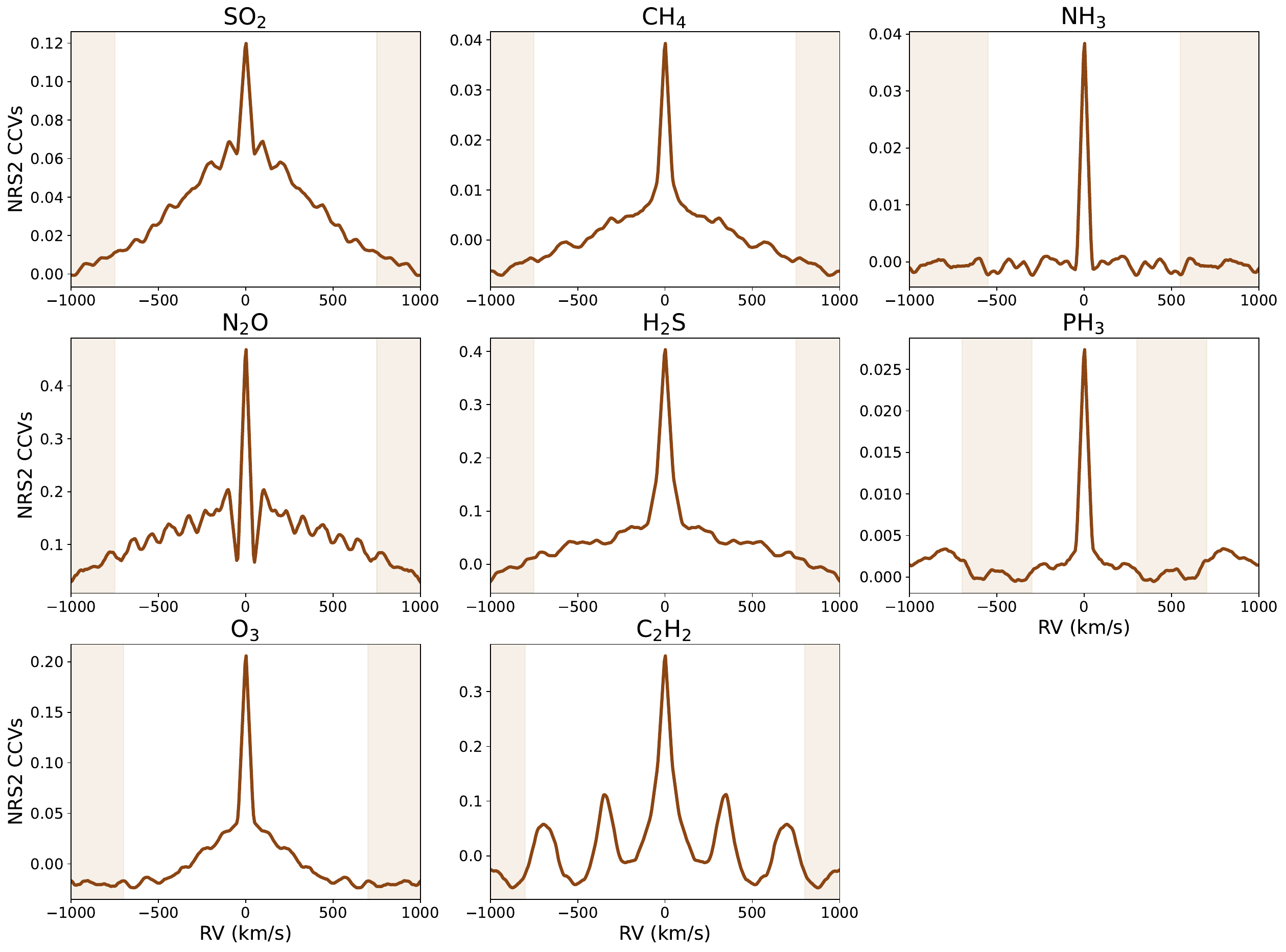}

\caption{Definition of the baseline regions (shaded regions) with respect to the template-template CCF within the optimal wavelength ranges using the Gaussian normalization on NRS2 for the non-detected molecules.
}
\label{fig:Freq_Baselines_nondetected_NRS2}
\end{figure*}

\newpage

\section{Results of the systematic cross-correlation search throughout NRS1 and NRS2 detectors.} \label{ap: sys_results}

\begin{figure*}
\includegraphics[width=\textwidth]{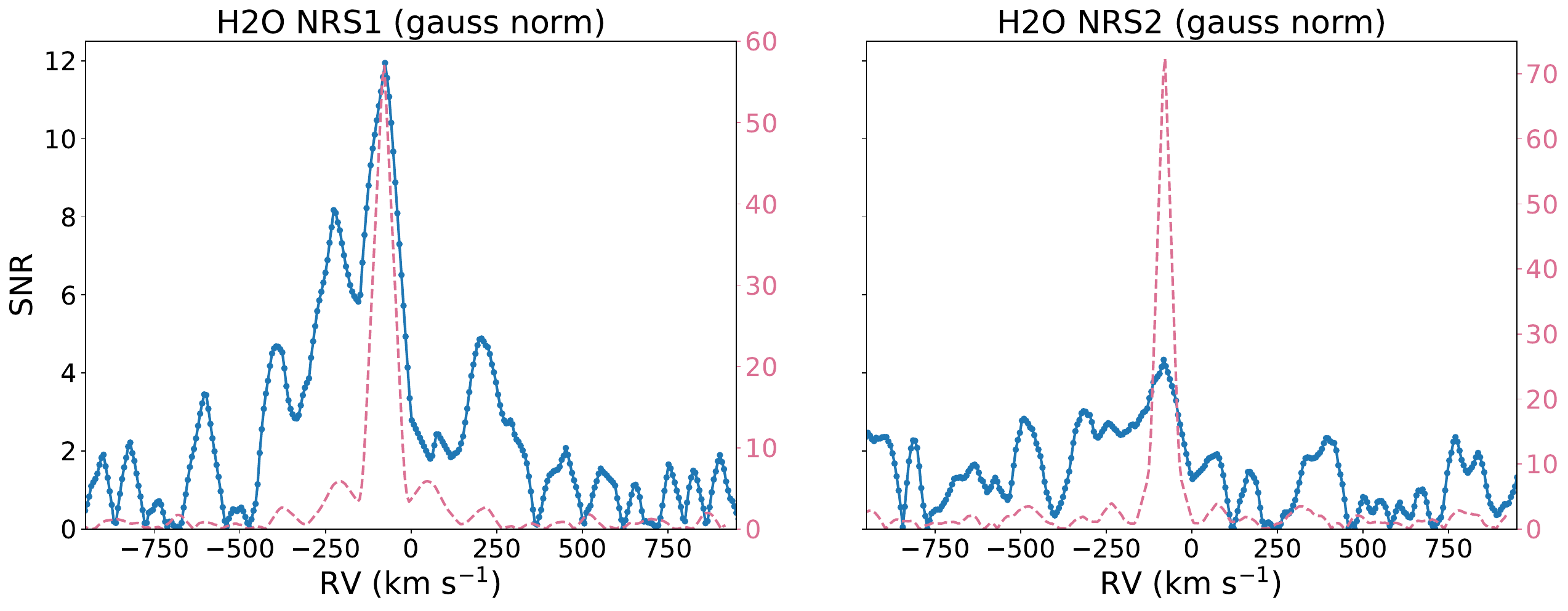}
\includegraphics[width=\textwidth]{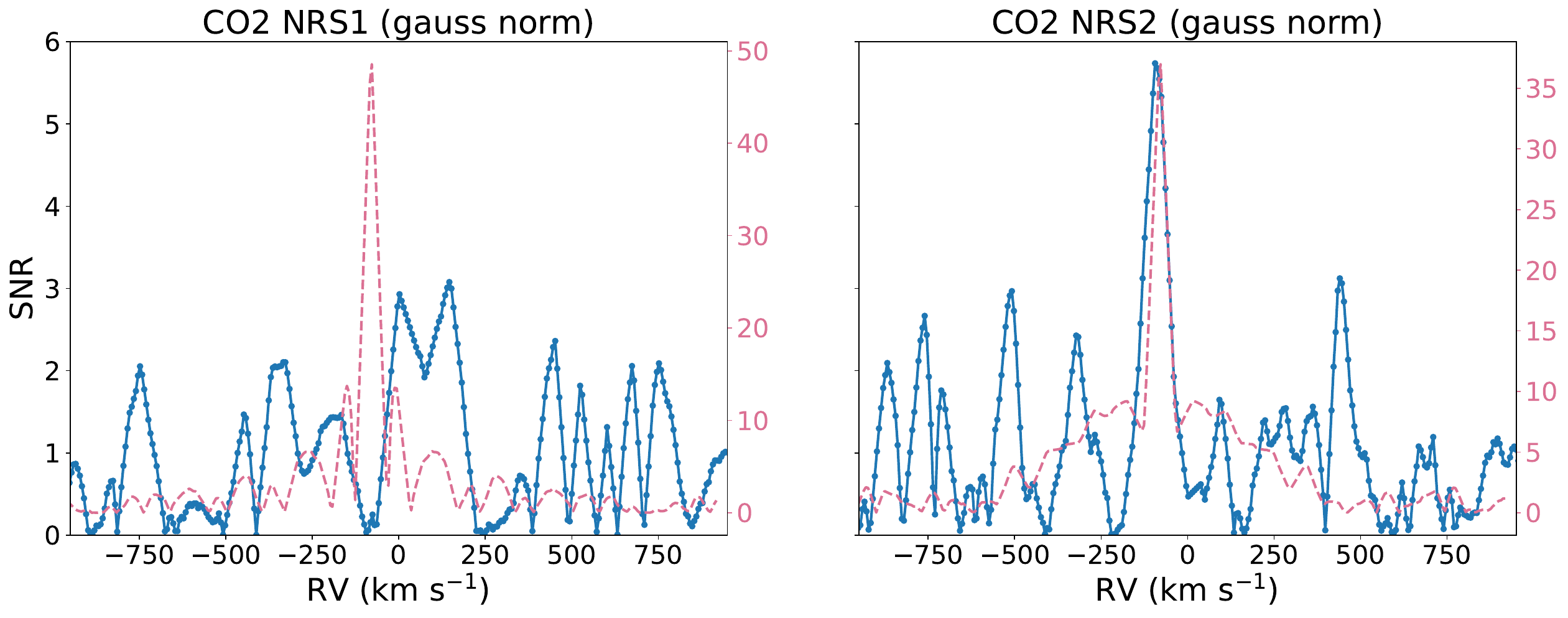}
\includegraphics[width=\textwidth]{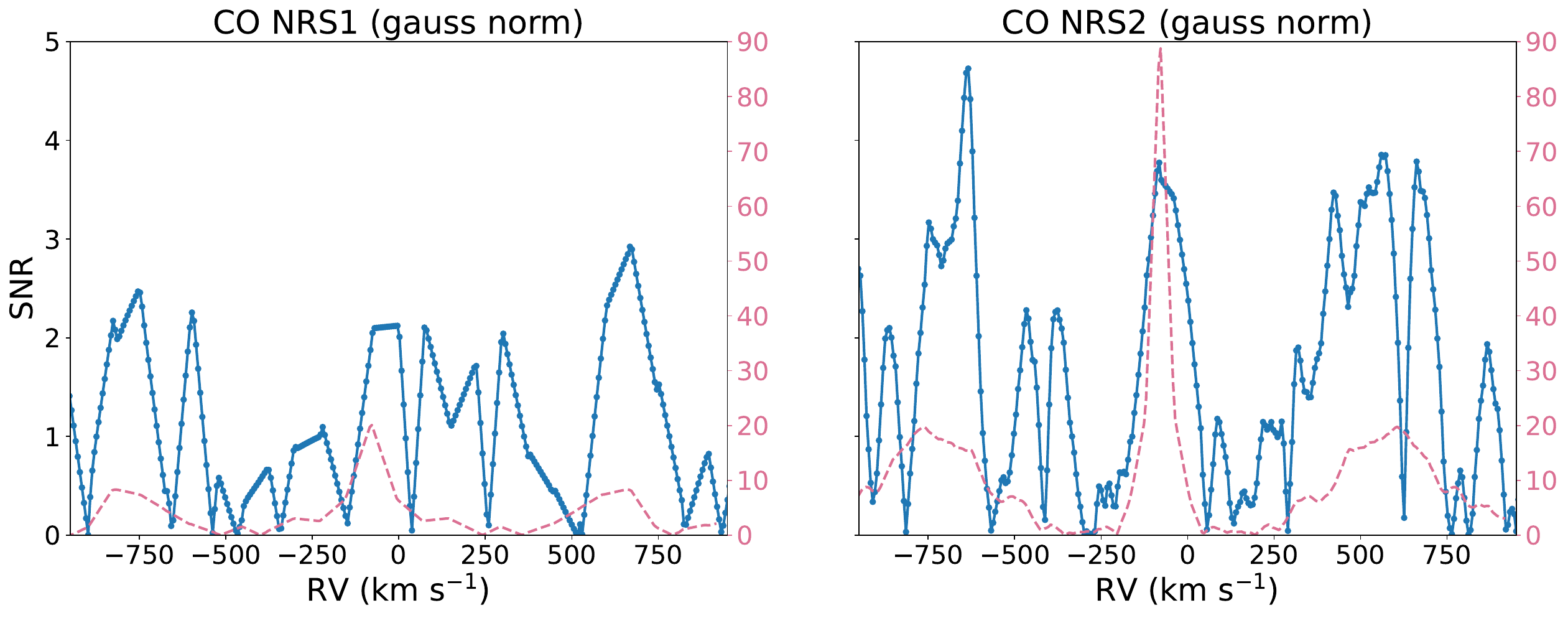}
\caption{Results of the systematic cross-correlation search of $\rm H_{2}O$, $\rm CO_{2}$ and $\rm CO$ along the whole wavelength range of NRS1 and NRS2 detectors using Gaussian normalization. The blue dash-dotted line shows the data-template CCF and the pink dashed line shows the template-template CCF.
}
\label{fig:CCFs_Gauss}
\end{figure*}

\begin{figure*}
\includegraphics[width=\textwidth]{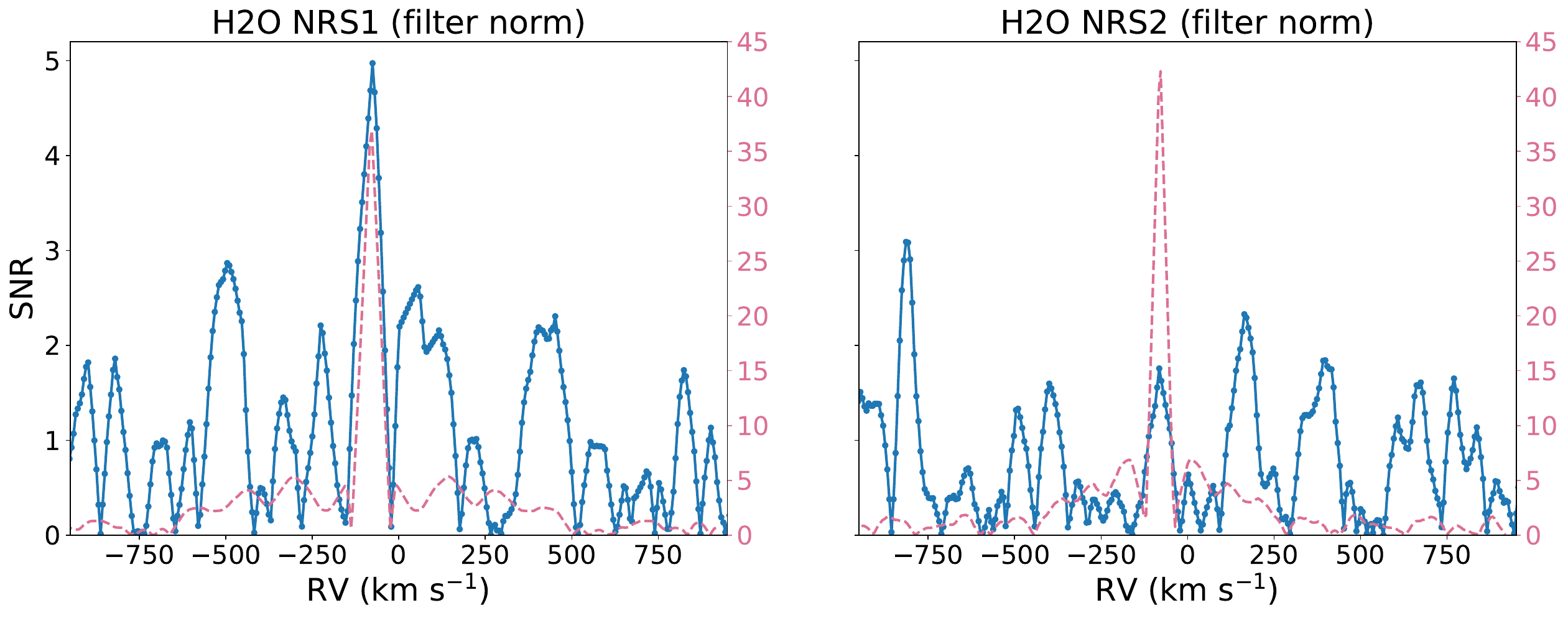}
\includegraphics[width=\textwidth]{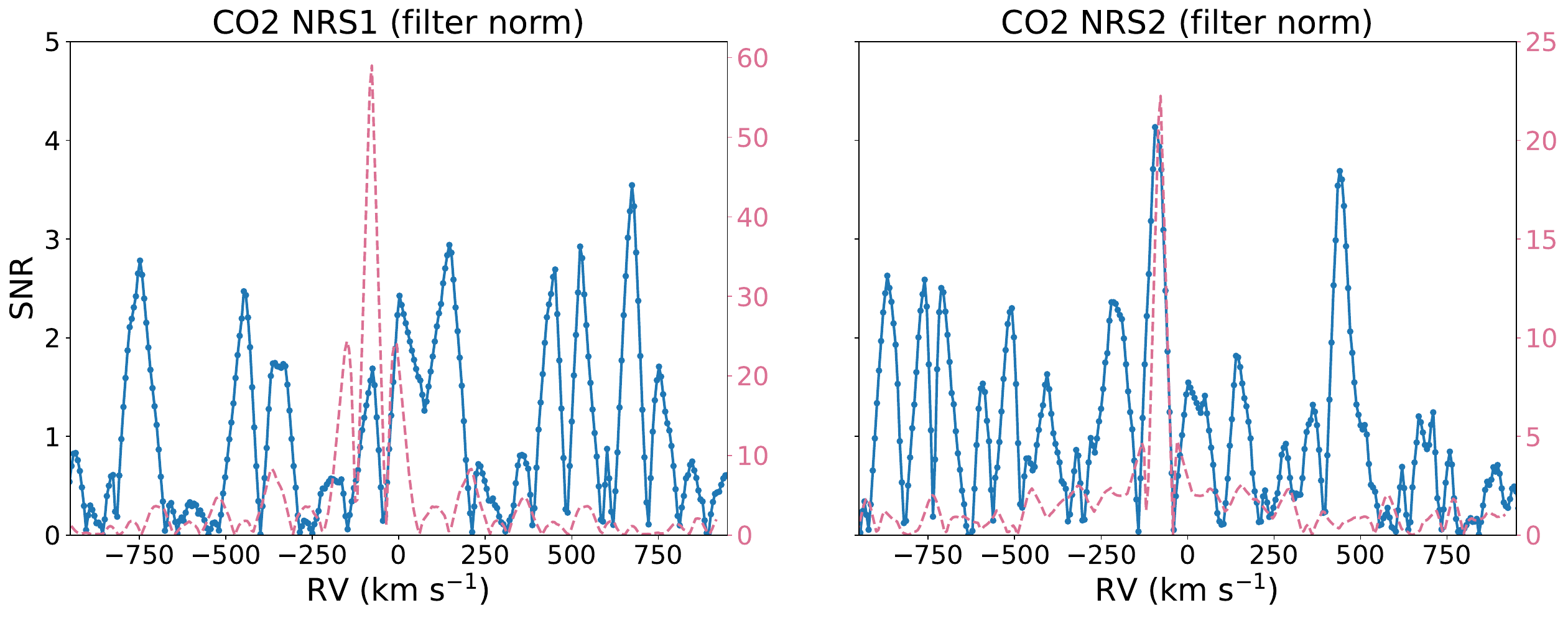}
\includegraphics[width=\textwidth]{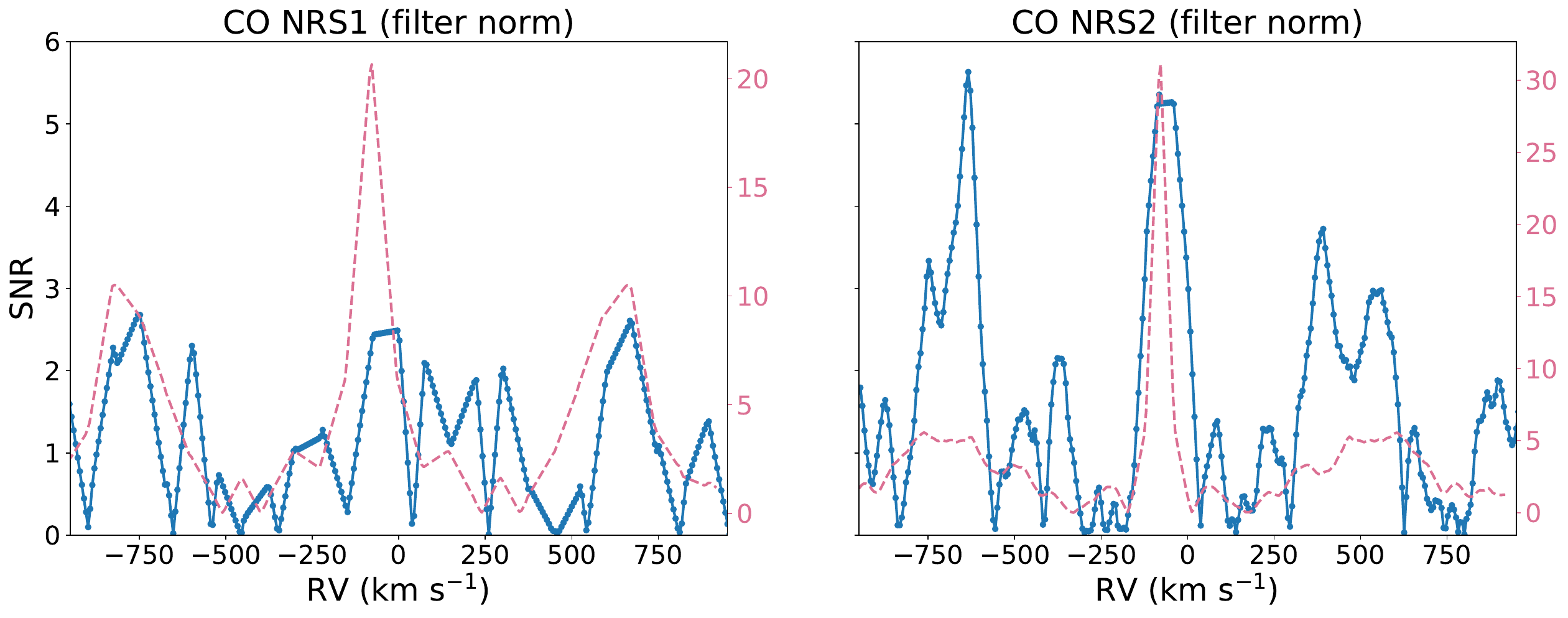}
\caption{Results of the systematic cross-correlation search of $\rm H_{2}O$, $\rm CO_{2}$ and $\rm CO$ along the whole wavelength range of NRS1 and NRS2 detectors using frequency-filter normalization. The blue dash-dotted line shows the data-template CCF and the pink dashed line shows the template-template CCF.
}
\label{fig:CCFs_freq}
\end{figure*}

\begin{figure*}
\includegraphics[width=0.45\textwidth]{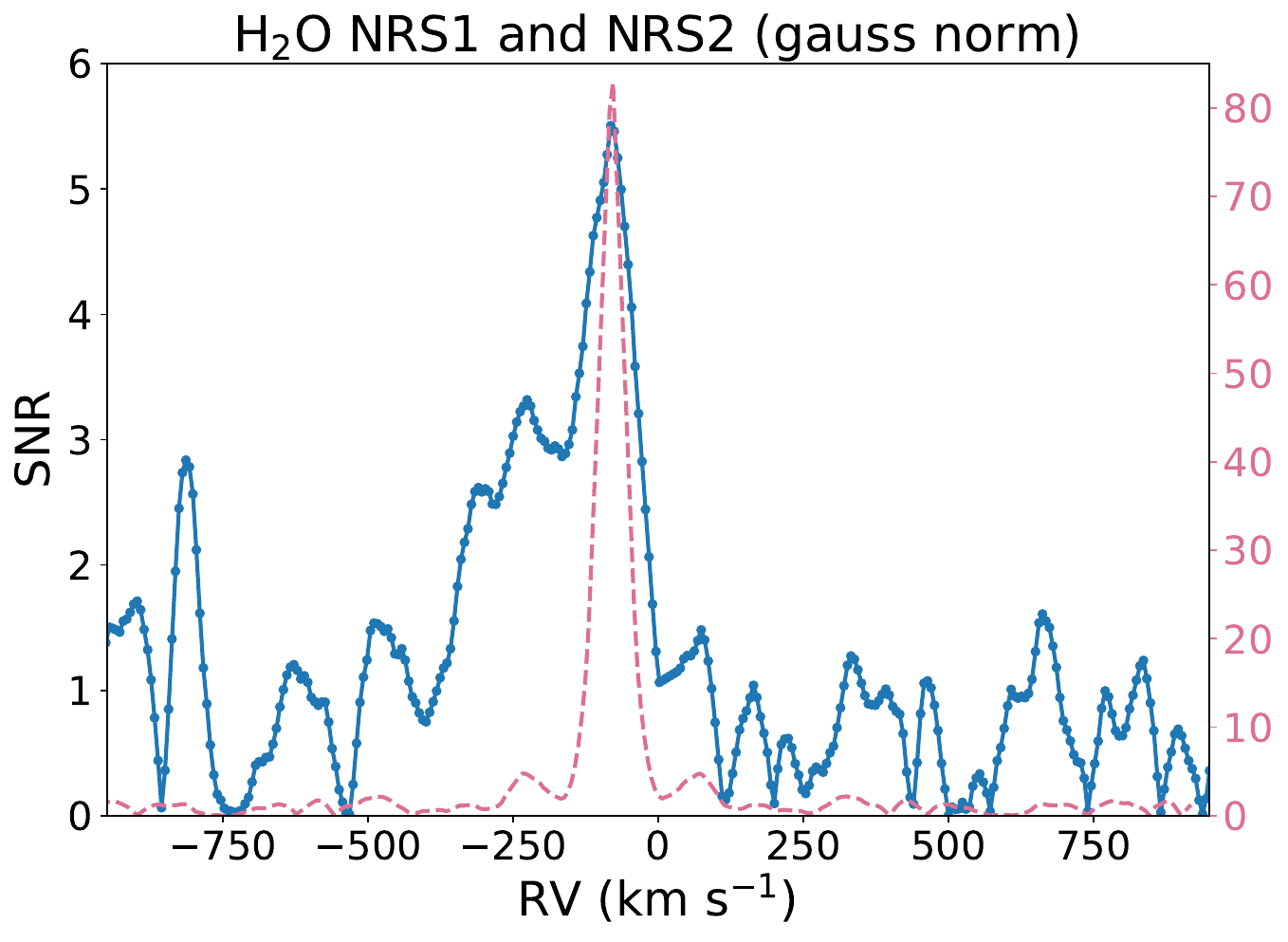}
\includegraphics[width=0.45\textwidth]{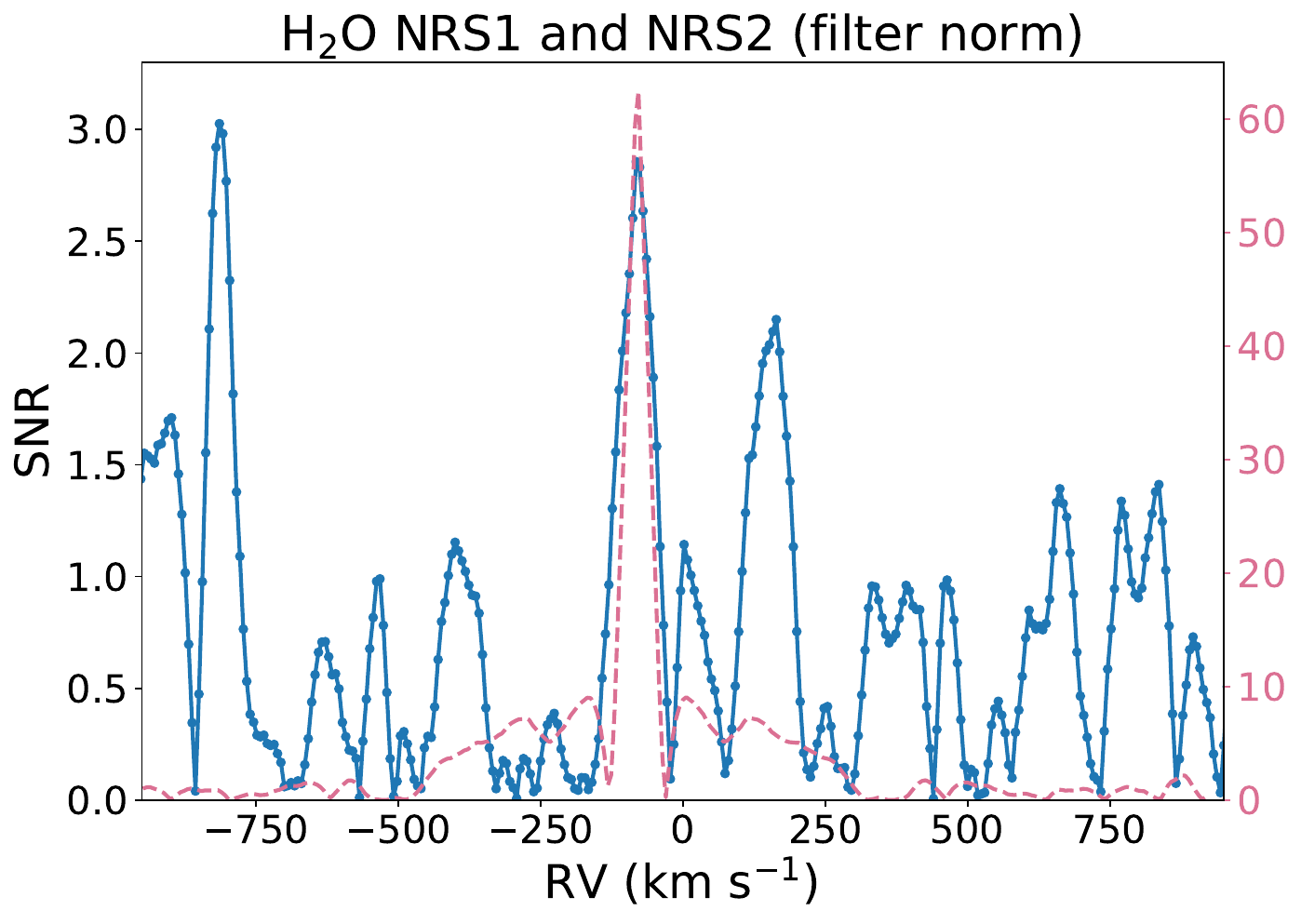}
\includegraphics[width=0.45\textwidth]{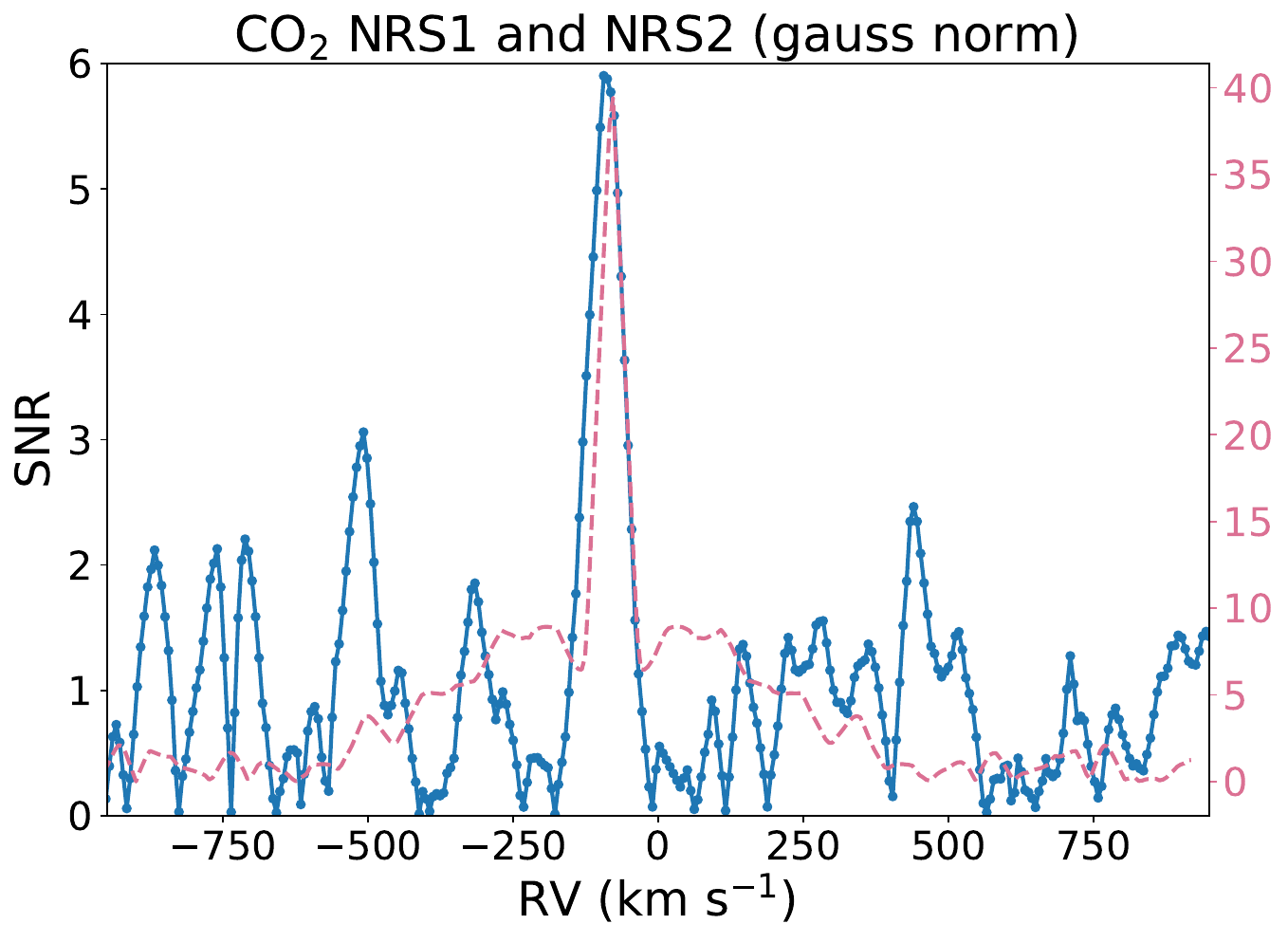}
\includegraphics[width=0.45\textwidth]{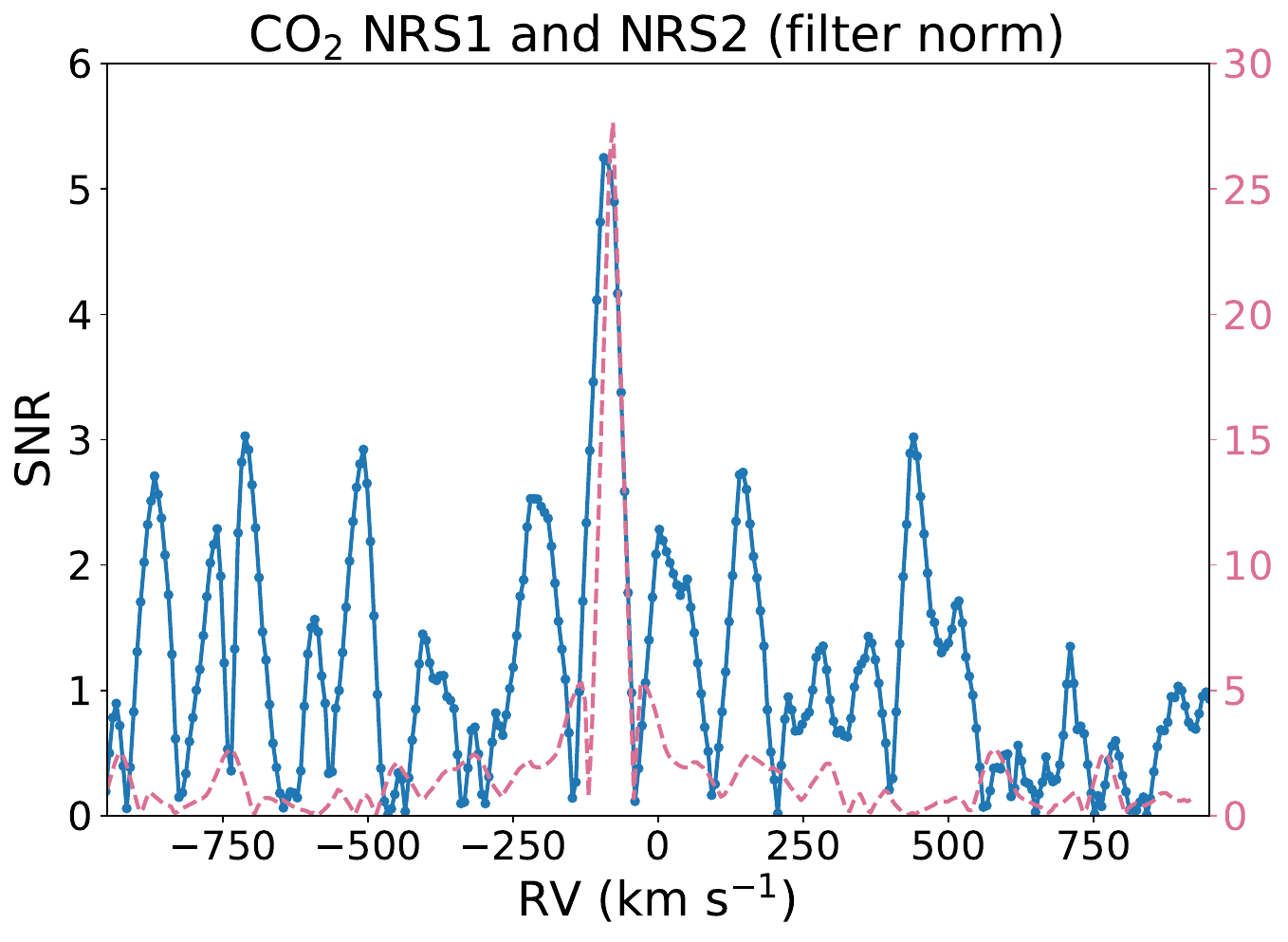}
\includegraphics[width=0.45\textwidth]{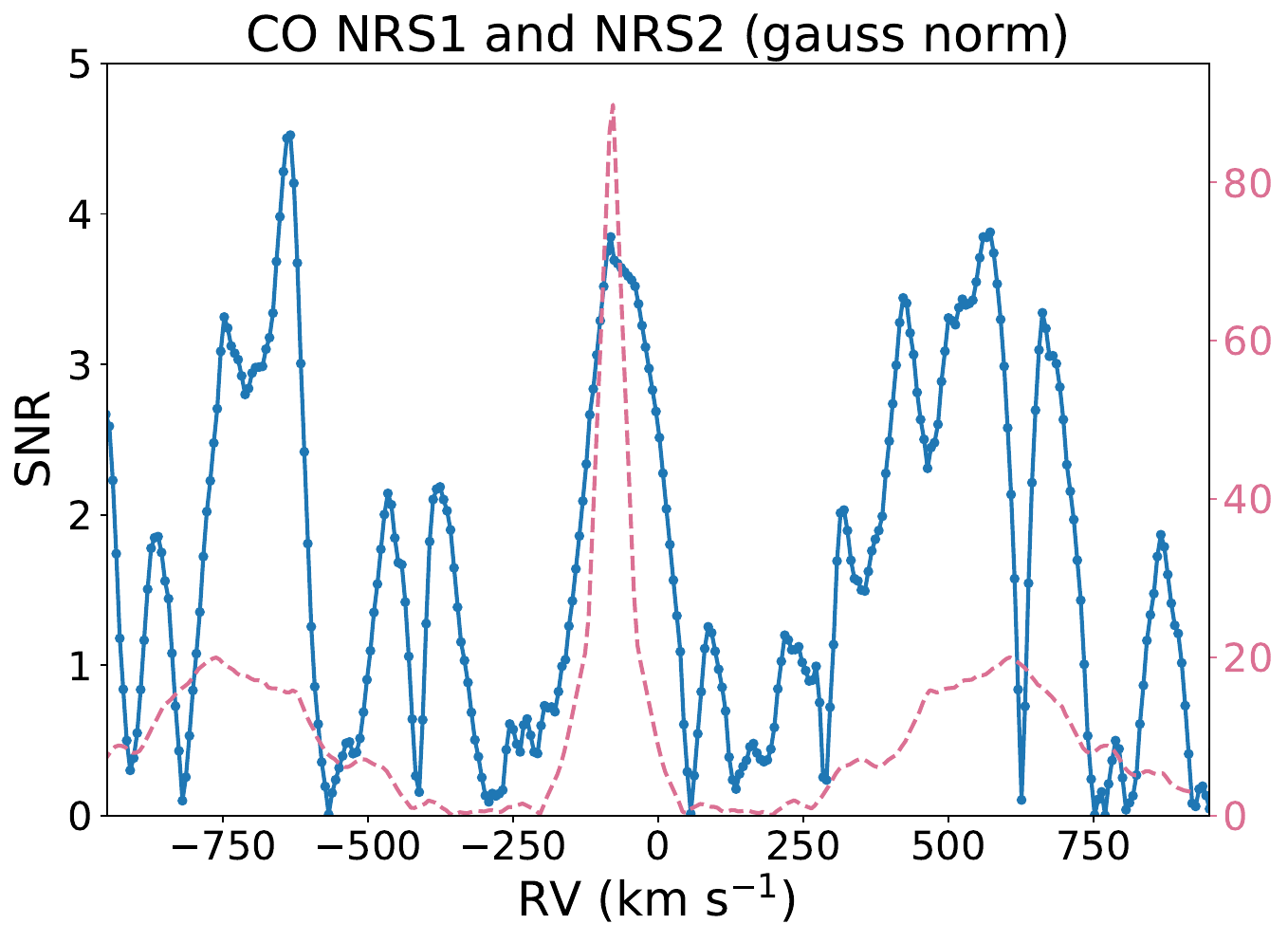}
\includegraphics[width=0.45\textwidth]{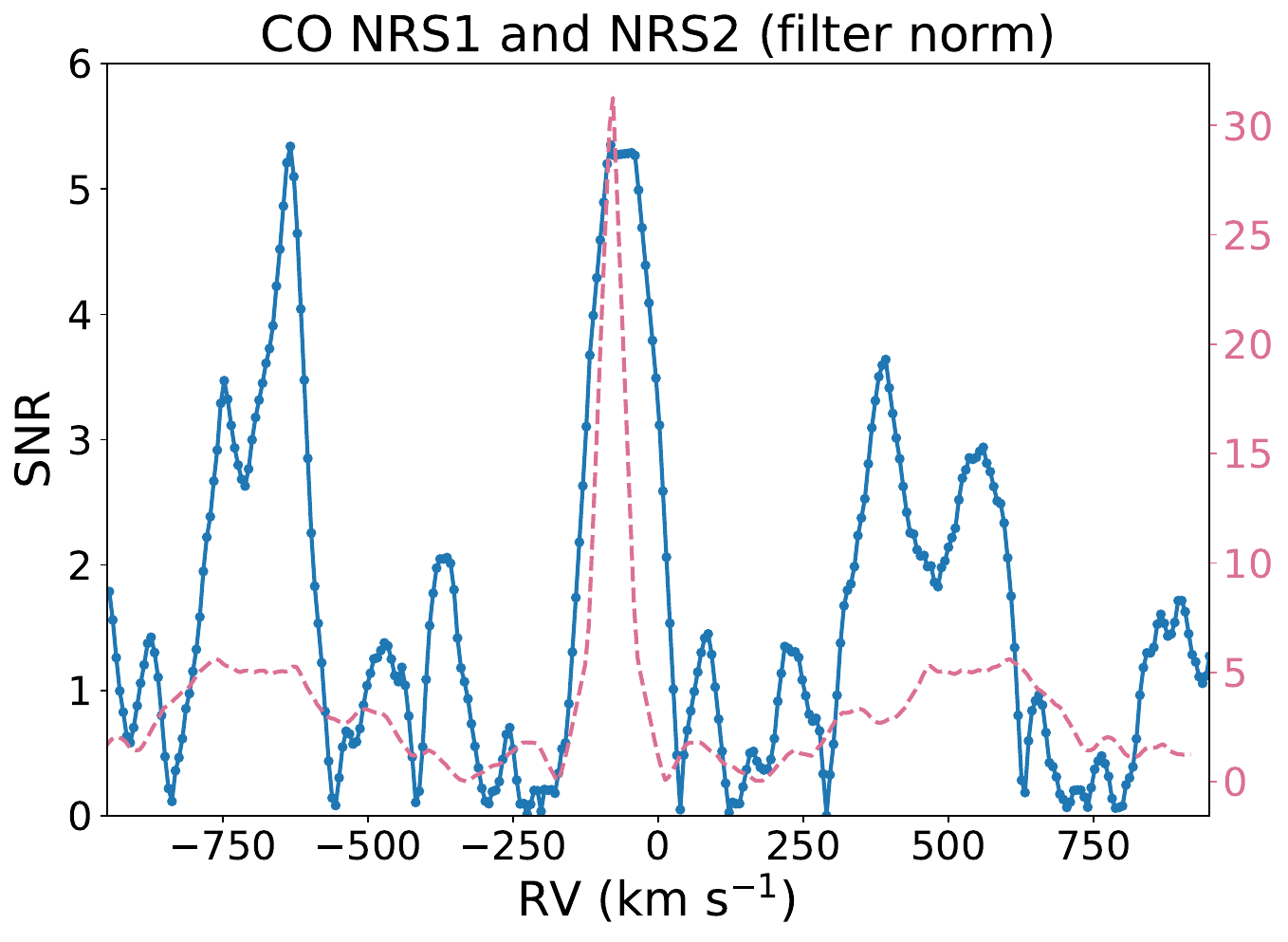}
\caption{Results of the systematic cross-correlation search of $\rm H_{2}O$, $\rm CO_{2}$ and $\rm CO$ along the whole wavelength range of NRS1 and NRS2 detectors using both normalizations. The blue dash-dotted line shows the data-template CCF and the pink dashed line shows the template-template CCF.
}
\label{fig:CCFs_CombNRS12}
\end{figure*}


\bsp	
\label{lastpage}
\end{document}